# Basic ideas of microscopic physics of liquid helium-4 and λ-transition to a coherent state

S.Yv. Morozov[*,1]

*" Institute for Physics and Power Engineering A.I. Leypunsky"*
*249033, Kaluga region, Obninsk, Russia, [*]leading researcher in retiring*

**Abstract**

The article formulates the basic principles of microscopic physics of a macroscopic ensemble of interacting helium-4 atoms. The concept of a vortex field is introduced, which is caused by the motion of the particles of the electron and nuclear subsystems of atoms of the macrosystem relative to each other. It is assumed that helium-4 atoms "immersed" in the internal vortex field acquire the properties of fermionic particles due to the atomic pseudospins $s = 1/2$ generated by the vortex. On this basis, a heuristic evolutionary model of helium-4 as an ensemble of atom-like quasiparticles with fermionic properties is constructed. As the temperature decreases, bonds are formed between pseudofermionic quasiparticles by pairing their pseudospins. The formation of a hierarchy of composite quasiparticles on this basis leads to the transition of helium-4 atoms from the gas phase to the liquid state, then to the state of incoherent Bose condensate, and finally to the state with a long-range order of atom-like quasiparticles. The structural forms of composite quasiparticles in different phases, the mechanisms of their formation, and the temperatures of these phase transitions ($T_{cr}$=5.22$K$, $T_0$=4.14$K$ and $T_\lambda$=2.175$K$, respectively) are determined. The ideas and calculations presented in this paper form the basis for solving the problem of the temperature dependence of the heat capacity of liquid helium-4 in both the helium-I and helium-II phases. The obtained solution has a logarithmic dependence of the heat capacity on the reduced temperature ($|1-T/T_\lambda|$) in the region of the λ-point of the phase transition of helium-4 from a disordered to an ordered state. The numerical calculations of the heat capacity in the vicinity of the temperature $T_\lambda$ are in very good agreement with the experimental data.

---

[1] *e-mail: starshoy.frost@gmail.com*

**Contents**

**Foreword**

The original article is written in Russian and translated into English by the author, who begs readers pardon for the poor English. English text is presented first on the 29 pages. This is followed by Russian text in the 30 pages. Both texts have their own page numbering.

**Introduction**

In 1932year W. H. Keesom and his co-workers discovered an anomalous increase in the heat capacity of liquid helium-4 when approaching the temperature $T_\lambda \approx 2.175K$ [1, 2]. He called this temperature the λ-point, the states of liquid helium above and below the $T_\lambda$ temperature were called the "helium-I" and "helium-II" phases, respectively, and the phase transition itself was called the λ-transition. Further studies have found that in phase II helium demonstrates unique properties of a coherent quantum system, which are manifested, in particular, in the effect of dissipative flow of liquid through capillaries [3, 4]. Understanding the nature of superfluidity is impossible without understanding the physics of liquid helium at the microscopic level. The solution to this problem is of fundamental interest for the physics of all known cooperative quantum phenomena.

The problem of the collective behavior of helium-4 atoms in the liquid phase is solved in this work by representing this system as an ensemble of quasiparticles with the properties of fermions. This representation means that the interaction between atoms can be expressed through the effect of the gaining of the pseudospin degrees of freedom with a half-integer value of their own mechanical moment by helium-4 atoms. This effect, in turn, leads to the" transmutation " of interacting boson atoms into free fermionic quasiparticles.

Self-organization of fermionic quasiparticles with decreasing temperature causes bosonization of the system due to the synthesis of new structural elements of the liquid in the form of atom-like composite quasiparticles with their successive complication and, ultimately, a transition to a coherent ordered state. According to the ideas of this work, the synthesis of atom-like quasiparticles based on structural elements of the previous level is carried out through the mechanism of formation of pseudomolecular complexes of atomic particles. The formation of these complexes in the process of decreasing temperature of the liquid is accompanied by the relaxation of their local structure to a pseudoatomic state. In turn, the pseudoatomic state of quasiparticles is the initial one for the formation of pseudomolecular quasiparticles of the next level of complexity on their basis. The realization of this kind of "atomic-molecular" states determines the formation of internal degrees of freedom of composite formations corresponding to the motion of structural elements of atomic-molecular quasiparticles relative to each other. We will call these degrees of freedom "pseudoatom" degrees of freedom.

The logic of the formation of structural elements of a quantum liquid in this work is based on the similarity of the principles and mechanisms of the organization of composite quasiparticles of helium-4 with the principles and mechanisms of the formation of the outer electron shells of molecules and atoms. In what follows, for brevity, this principle of similarity of the mechanisms of organization of composite quasiparticles of a liquid by the mechanisms of formation of molecules and atoms we will call "*the principle of AM-similarity of the structure of a liquid*".[2].

---

[2] In this regard, when describing the mechanisms of establishing bonds between the structural elements of a liquid during the formation of composite quasiparticles, we will formally use the terminology and designations that are used to describe and designate atomic the electronic states of atoms and interatomic bonds in the physics of atoms and molecules.

In the first part of the work, the main ideas of the microscopic physics of liquid helium are presented[3]. In the second part of the work, these ideas are used to calculate the heat capacity of helium-4 in the vicinity of the $\lambda$-transition of the system to a coherent state.

## Part 1. Microscopic physics of the generation of fermionic properties of a macroscopic ensemble of helium-4 atoms and their evolution with decreasing temperature.

*"Very often such, a simplified model throws more light on the real workings of nature than any number of "ab initio" calculations of individual situations…"*

***Philip W. Anderson, Nobel Lecture, 1977***

### 1.1 Interacting atoms of helium-4 as fermionic pseudoatoms. Gyrotons.

#### 1.1.1 *Vortex field and induced spins of atoms. Formation of atom-like quasiparticles.*

The effect of interatomic interaction of a macroscopic ensemble $N_0$ of helium atoms in a volume $V$ is associated in this paper with the formation of an internal vortex field created by electron and nuclear charges of the macrosystem moving relative to each other. From this point of view, the helium atoms are "immersed" in an internal vortex field, which determines the effect of transmutation of an ensemble of interacting boson atoms into a system of free fermionic quasiparticles.

The area of space surrounding each atom, the volume of which is equal to the specific volume of the macrosystem $\upsilon = V/N_0$, can be considered as a "unit cell" of the quasi-stationary localization of the atom. The response of an atom to a vortex field depends on the temperature of the system and in general can be characterized by two mutually dependent effects: first, the vibration of the electron orbitals of atomic electrons and, second, the quasi-closed motion (oscillations) of the helium atom in the vicinity of the center of the "unit cell". Herewith the second effect can manifest itself explicitly only in the condensed state of the substance.

From the point of view of the semiclassical representation, the vibration of electronic orbitals is an unsteady precession of the planes of orbital motion of atomic electrons. This form of "additional" motion of the electron subsystem of an atom in a vortex field can be regarded as an effect that generates an "*imaginary electron*" localized on the atom. The creation of an imaginary electron is the result of the excitation of the electron subsystem of the atom. As a consequence, the statistics of the realized quantum states of particles caused by such excitations should correspond to the statistics of fermions with spin determined by the quantum number $s=1/2$ (in the future, the value of the angular momentum will be determined by the value of the corresponding quantum number). In this case, an imaginary electron localized on a neutral helium atom causes the helium atom to "acquire" a new quality, which we define as "*induced spin*" (or "*pseudospin*") with angular momentum $s = 1/2$. And, thus, "transforms" a boson into a fermionic quasiparticle. From this point of view, the imaginary electron is the carrier of the pseudospin state of the atom, due to the interaction of the atom with the vortex field. A helium atom with an induced spin localized on it is a fermionic quasiparticle, which we will call a "*gyroton*". Thus, defining the vortex field as an intrinsic property of an atomic particle, we proceed from analyzing the behavior of an ensemble of interacting helium-4 atoms to considering an ensemble of "free" fermion quasiparticles.

---

[3] To reduce the volume of the article and to restrict the number of newly introduced concepts, in this paper we will simplify or omit those positions and details of the microscopic structure of quasiparticles that are not used in the calculation of the heat capacity in the vicinity of $\lambda$-transition in an explicit form.

According to the above, the gyroton in the temperature range $T<T_{cr}$, where $T_{cr}$ corresponds to the critical temperature of the transition of helium-4 from a gaseous to a liquid state, performs a quasi-closed motion in the vicinity of the center of its quasi-stationary localization (except for other motions with much longer characteristic times, which we do not consider). Such a dynamic formation, contained in the corresponding "unit cell", can be regarded as a gyrotonic pseudoatom. We will call this atom-like fermionic quasiparticle a "*gyroatom*". In this representation, gyrotons play the role of pseudoatomic electrons with an effective mass equal to the mass of the helium atom $M_{He}$. Therefore, gyrotons in the appropriate cases will be called pseudoelectrons. In this case, the gyroatoms themselves have both pseudospin and pseudo-atomic degrees of freedom[4].

The behavior of fermionic gyrotons in a medium of similar particles depends on temperature and is determined by the structure and quantum state of the composite gyroatoms (atomic-molecular quasiparticles) formed on their basis. The formation of atomic-molecular quasiparticles determines the fundamental possibility of separating their motion into pseudoatomic and pseudospin degrees of freedom.

### 1.1.2 *Requirements and form of realization of fermionic properties of boson atoms.*

The Bose-Fermi mechanism of transmutation of the statistical properties of helium-4 atoms is associated with the restriction of the freedom of their translational motion in the field $\Phi$ of potential interatomic interaction. As a result, part of the thermal energy of translational motion of atoms is "converted" into the intra-atomic energy of the pseudospin degrees of freedom of gyroton quasiparticles. Herewith, the energy of all "realized" pseudospin degrees of freedom turns out to be completely "compensated" by the negative energy of the interaction of atoms, which is "spent" on changing the nature of the motion of atoms. That is, for $N_a$ atoms to "acquire" spin degrees of freedom and, thus, for the formation of $N_e=N_a$ gyroton quasiparticles, it is necessary that the negative potential energy of the interaction of all $N_0$ atoms of the ensemble in absolute value should be equal to the energy of the thermal translational motion of $N_a$ degrees of freedom of atoms at temperature $T$:

$$|\varphi(\upsilon)| = (N_e/N_0) \cdot (1/2) k_B T \qquad (1)$$

here $\upsilon = \upsilon(T)$ is the temperature-dependent specific volume of the atomic system; $|\varphi(\upsilon)|$ is the modulus of the potential energy of the interaction of helium atoms per particle.

From the standpoint of representing helium as an ensemble of pseudoatomic quasiparticles and in accordance with the previously mentioned "the principle of AM-similarity of the structure of a liquid" (see Introduction, p. 4), the induced spins "generate" gyroatomic pseudoelectrons, which occupy the 1*s* state of the pseudoelectron shell of the corresponding gyroatom. The interaction between atoms leads to the splitting of the 1*s* level of the pseudoelectron shell and the formation of a band of energy states of gyrotons. The energy band width is equal to the energy $|\varphi(\upsilon)|$ and increases with the number of induced imaginary electrons as the temperature of the system decreases. Herewith, according to (1), the internal energy of the pseudospin states of atoms completely compensates for the energy of the interatomic interaction until all possible imaginary electrons are induced at a certain characteristic temperature $T_0$. Therefore, we may accept that the energy states of the gyroatomic pseudoelectrons correspond to the energy levels of an absolutely degenerate system of free fermions with an effective temperature $T_{eff}=0K$, a mass $M=M_{He}$, and a spin $s = 1/2$. And this condition remains until the modulus of negative energy of the interaction of atoms reaches its maximum value $|\varphi_0|$.

---

[4] In the further presentation, we will sometimes omit the prefix "pseudo" or "quasi" in the words in cases when it will be obvious that we are talking about the structural-dynamic elements of gyroatomic quasiparticles.

The total number of the imaginary electrons $N_{e0}$ that can be realized in a system of $N_0$ helium atoms is equal to the number of atomic electrons $2N_0$. It follows that when the system temperature decreases, the process of generating induced spins and, therefore, the formation of gyroatomic pseudoelectrons, includes two stages. The first stage corresponds to the process of the generation of $N_e=N_0$ imaginary electrons, when all $N_0$ particles of the atomic ensemble "acquire" pseudospin degrees of freedom in the vortex field and, thereby, are transformed into fermionic quasiparticles. Herewith, the movement of each helium atom along one of the three translational degrees of freedom turns out to be "frozen". From the point of view of the ideas of this work, this corresponds to the requirements that the system reaches the critical thermodynamic parameters $T_{cr}$, $V_{cr}$ and $P_{cr}$ of the transition of the system from the gaseous to the liquid state. The corresponding pseudoelectrons will be called primary pseudoelectrons or $K$-pseudoelectrons. $K$-pseudoelectrons are realized at $T \geq T_{cr}$ and gradually fill the states of the $K$-band as the temperature decreases. These states correspond to the ground state of free fermions with $T_{eff}=0K$. The total number of $K$-pseudoelectrons at a temperature $T=T_{cr}$ is equal to $N_{K0}=(1/2)N_{e0}=N_0$ and all the atoms of the system get the properties of fermions. According to requirement (1), the boundary energy of $K$-pseudoelectrons at $T=T_{cr}$ corresponds to the Fermi energy $E_F(N_{K0})=k_BT_{cr}/2$:

$$E_{F_{cr}}(N_{K0}) = \frac{1}{2}k_BT_{cr} = \frac{\hbar^2}{2M_{He}} \cdot \left(\frac{6\pi^2 N_{K0}}{g_s \cdot V_{cr}}\right)^{2/3} = \frac{\hbar^2}{2M_{He}} \cdot \left(\frac{3\pi^2}{\upsilon_{cr}}\right)^{2/3} \qquad (2)$$

Here, the degeneracy factor $g_s=2s+1$ corresponds to the spin state of pseudoelectrons $s=1/2$; $V_{cr}=V(T_{cr})$ and $\upsilon_{cr}=V_{cr}/N_0$ are the critical volume and specific critical volume of helium-4 at $T=T_{cr}$.

.

## 1.2 The first stage of condensation of quasiparticles in helium. Condensation of primary $K$-pseudoelectrons and formation of fermi-liquid of two-particle gyroatoms.

### 1.2.1 *The mechanism of formation of pair-correlated states of gyroatoms and the formation of a subsystem of "secondary" pseudoelectrons.*

From the point of view of the quasiparticle representation, only half of the maximum possible number of $N_{e0}$ states of the band of pseudoelectronic quasiparticles are filled at the temperature $T=T_{cr}$ (see Subsection 1.1.2). The modulus of the potential energy of interaction of atoms per particle at $T<T_{cr}$ becomes greater than the energy of internal motion of independent spin degrees of freedom of primary pseudoelectrons. At the same time, the possibilities of one independent atom to compensate for the "excess" potential energy by creating a free pseudospin state are exhausted. Therefore, the system becomes unstable with respect to the formation of new structural units of the macrosystem by establishing pair-correlation relationships between gyroatoms.

The mechanism of establishing pair correlations of gyroatoms can be associated with the process of $s$-pairing of the spins of those $K$-pseudoelectrons that are on the Fermi surface at the current temperature $T<T_{cr}$. This means the "condensation" of $Ks$-pseudoelectron states on the Fermi surface of unpaired $K$-pseudoelectrons. Herewith, paired $K$-pseudoelectrons lose the property of individual fermionic quasiparticles. This process is a manifestation of the effect of the phase transition of helium from a gaseous to a liquid state.

The gyroatomic bonds caused by the pairing of $K$-pseudoelectrons will be called primary bonds, and the pair-correlated states of fermionic gyrotons will be called "*primary pairs*". A primary pair of gyroatomic quasiparticles is a virtual formation, which is a manifestation of the existence of a certain short-range order parameter characterizing a monoatomic macrosystem in a liquid aggregate state.

The specific volume $\upsilon(T)$ of the liquid decreases and the degree of restriction of the freedom of translational motion of individual atoms increases with a decrease in temperature $T<T_{cr}$ and the establishment of primary pair bonds. As a consequence, the conditions for the second stage of the process of generation of imaginary electrons and, thereby, for the formation of "secondary" pseudoelectrons are realized. According to the previous section, this process occurs in the temperature range $T_0 \leq T < T_{cr}$.

According to the above, the condensation of $K$-pseudoelectrons is accompanied by the generation of secondary induced spins of helium atoms up to the temperature $T_0$. At the same time, the total number of primary and secondary free pseudoelectrons in the temperature range $T_0 \leq T < T_{cr}$ remains equal to $N_0$. Therefore, each gyroatom participating in the pair-correlation interaction carries one unpaired secondary pseudoelectron realized at $T<T_{cr}$. That is, helium atoms, as individual quasiparticles, retain the status of gyrotonic fermions in the temperature range $T_0 \leq T < T_{cr}$. For particles forming primary pairs, this status is provided by secondary induced spins. Herewith, the energy states of the secondary pseudoelectrons, as in the case of the realization of $K$-pseudoelectrons, correspond to the states of fermions with an effective temperature $T_{eff} = 0K$, a mass $M = M_{He}$, and a spin $s=1/2$.

Thus, during the transition of the system to a liquid state, on the one hand, primary pairs of gyroatoms are formed. On the other hand, in the same temperature range $T_0 \leq T < T_{cr}$, secondary gyrotonic pseudoelectrons are formed, whose energy states correspond to the states of free fermions. Moreover, both processes occur synchronously. From this we can conclude that the secondary pseudoelectrons are realized in pairs and, accordingly, fill the states of an absolutely degenerate system of fermions inside the Fermi sphere in pairs. This means that the secondary pseudoelectrons of the primary pairs have oppositely directed spins. The presence of paired correlations between particles carrying free spins of an absolutely degenerate fermion system can be considered as an effect of the formation of the ground state of a gyroton Fermi liquid.

The twofold nature of pair-correlated gyroatoms at $T<T_{cr}$ determines the formation of common pseudoatomic and pseudospin degrees of freedom in them. This allows us to consider the primary pairs of gyroatoms as two-particle pseudoatoms, which we will call "*bigyrotons*". Note that there is no explicit separation of the motion of gyroatoms according to the pseudoatomic and pseudo-spin degrees of freedom at the stage of the formation of the fermionic properties of the atomic system, since this requires that the interaction energy of atoms be greater than the kinetic energy of pseudoelectronic states of gyrotons. Therefore, up to the temperature $T_0$ of the completion of the formation of the subsystem of secondary pseudoelectrons, their energy states correspond to the states of free fermions.

From the standpoint of the principle of AM-similarity of the structure of a liquid (see Introduction, p. 4), secondary pseudoelectrons of primary pairs occupy states in the $L$-pseudo-shell of bigyroton pseudoatoms. Therefore, the secondary pseudoelectrons will be called "$L$-pseudoelectrons". And these $L$-pseudoelectrons fill in pairs the states with opposite spins in their "own" (secondary) $L$-band, the energy levels of which correspond to the energy levels of an absolutely degenerate system of fermions. The total number of pseudoelectron states realized at the current temperature $T_0 \leq T < T_{cr}$ is $N_e = N_{K0}+N_L$. In this case, the number $N_L$ of created $L$- pseudoelectron quasiparticles is obviously equal to the number of $N_{K_-}$ condensed $K$-pseudoelectrons. The energy of the $L$-pseudoelectrons filling the $L$-band at the current temperature is measured from the energy of the upper filled level of the $K$-band, taken as a zero. At the temperature $T<T_{cr}$, this band is filled with $N_K=(N_{K0}-N_{K_-})=(N_0-N_L)$ unpaired $K$-pseudoelectrons. Thus, $L$-pseudoelectrons at temperature $T$ fill the states of free fermions with $T_{eff}=0K$ and a chemical potential $\mu(N_L)_{T \geq T_0}$ equal to the Fermi energy $E_{F_L}(N_L)$:

$$E_{F_L}(N_L) = \frac{N_e}{N_0} \cdot \frac{1}{2} k_B T - \frac{(N_0 - N_L)}{N_0} \cdot \frac{1}{2} k_B T = \frac{N_L}{N_0} k_B T = \frac{\hbar^2}{2M_{He}} \cdot \left( \frac{6\pi^2}{g_s} \cdot \frac{N_L/N_0}{\upsilon(N_e)} \right)^{2/3} \quad (3)$$

Here, the specific volume of the pseudoelectronic system $\upsilon(N_e)$ is determined by the total number of realized pseudoelectronic states $N_e$ at a constant initial (in terms of the realization of total number pseudoelectrons) volume of the liquid $V_{cr}=V(T_{cr})$:

$$\upsilon(N_e) = \frac{V_{cr}}{N_e(T)} = \frac{\upsilon_{cr}}{1 + N_L/N_0} \quad (4)$$

The chemical potential $\mu_0$ and the corresponding Fermi energy $E_{F0}(N_{L0})$ of the system of $L$-pseudoelectrons at the temperature $T_0$ of realization of all $N_{L0}=N_0$ secondary pseudoelectrons, according to (1), (3) и (4), are equal to the modulus of the interatomic interaction energy per particle:

$$\mu_0 = E_{F0}(N_{L0}) = -\varphi_0 = k_B T_0 = \frac{\hbar^2 k_0^2}{2M_{He}} = \frac{\hbar^2}{2M_{He}} \cdot \left( \frac{3\pi^2 N_0}{V_0} \right)^{2/3} = \frac{\hbar^2}{2M_{He}} \cdot \left( \frac{3\pi^2}{\upsilon_0} \right)^{2/3} \quad (5)$$

Here $\varphi_0(\upsilon_0) = -k_B T_0$ corresponds to the maximum of the negative potential energy of pair interaction of atoms per particle (with an accuracy up to taking into account the internal energy of $Ks$-pairing of primary pairs); $V_0=V(T_0)$ is the volume of the atomic system at temperature $T_0$; $\upsilon(T_0) = V_0/N_0 = \upsilon_0$ corresponds to the specific volume of helium at $T=T_0$; $k_0$ is the boundary wave vector of the completely filled Fermi sphere with $L$-pseudoelectrons at $T=T_0$.

### 1.2.2 *Separation of states of a pseudoelectronic liquid into states of Ls - and Lp-type quasiparticles.*

According to Subsection 1.2.1, secondary pseudoelectrons fill the states of the $L$-shell of all bigyrotonic pseudoatoms at temperature $T_0$. In turn, the energy states of the $L$-shell are subdivided into states corresponding to the orbital angular momenta $l=0$ ($s$-subshell) and $l=1$ ($p$-subshell). Therefore, the splitting of the $L$-shell states, which occurs as a result of interatomic interaction, leads to the formation of an $L$-band consisting of $Ls$ and $Lp$ subbands. According to the previous subsection, the $L$-band at a temperature $T_0$ is filled with $N_0$ secondary pseudoelectrons with an effective temperature $T_{eff}=0K$ (eq.(5)). In this case, the $L$-pseudoelectrons inside the Fermi sphere of radius $k_0$ are divided into two subsystems (see Figure 1). The first subsystem includes $L$-pseudoelectrons with the energy of the relative motion of the atomic particles of the primary pair below the energy $\varepsilon_1$, determined by the orbital quantum number $l=1$. And, accordingly, the second subsystem is formed by $L$-pseudoelectrons with energy $\varepsilon \geq \varepsilon_1$. The corresponding subsystems of particles will be called the $Ls$ and $Lp$ components of pseudoelectrons. The quasiparticles of the two subsystems that fill corresponding subbands are denoted as $Ls$- and $Lp$-type particles.

On the one hand, all secondary pseudoelectrons at $T=T_0$ are free fermions with spins $s=1/2$. On the other hand, $Lp$-pseudoelectrons can be considered as an independent subsystem of pair-correlated quasiparticles, which have orbital angular momentum $l = 1$ (additional to the spin moment $s = 1/2$) . Herewith, $Lp$-quasiparticles of this kind pairs can be in states corresponding to quantum numbers $j = 1/2$ or $j = 3/2$ with equal probability. Then the degree of degeneracy of $Lp$ particles turns out to be two times greater than the degree of degeneracy of $Ls$ particles. In this case, the relative fractions of particles of $Ls$- and $Lp$- types can be found from the condition of equality of the specific thermodynamic potentials of two hypothetical fermionic subsystems with the degree of degeneracy of quantum states $g_{1/2}=2$ and $g_{3/2}=4$:

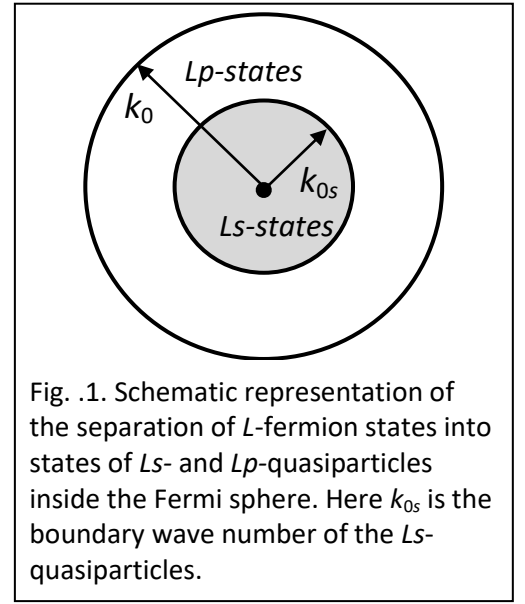

Fig. .1. Schematic representation of the separation of $L$-fermion states into states of $Ls$- and $Lp$-quasiparticles inside the Fermi sphere. Here $k_{0s}$ is the boundary wave number of the $Ls$-quasiparticles.

$$\frac{N_{0s}}{N_0} \cdot \frac{\hbar^2}{2M_{He}} \cdot \left(\frac{6\pi^2}{g_{1/2} \cdot \upsilon_0}\right)^{2/3} = \frac{N_{0p}}{N_0} \cdot \frac{\hbar^2}{2M_{He}} \cdot \left(\frac{6\pi^2}{g_{3/2} \cdot \upsilon_0}\right)^{2/3} \tag{6}$$

The $N_{0s}$ and $N_{0p}$ values correspond to the numbers of $Ls$ and $Lp$ particles, respectively, and $N_{0s}+N_{0p}=N_0$. The relative fractions of particles $N_{0s}/N_0$ and $N_{0p}/N_0$ determines the probability of finding fermionic quasiparticles in the $Ls$ and $Lp$ states, respectively.

From (6) we obtain that the relative fractions of $Ls$- and $Lp$- particles are equal[5]:

$$N_{0s} / N_0 = \left(1 + 2^{2/3}\right)^{-1} = 0.3865;$$
$$N_{0p} / N_0 = \left(1 - N_{0s} / N_0\right) = 0.6135 \tag{7}$$

The effects of the pair-correlated state of fermionic quasiparticles and their separation into $s$- and $p$- states are a manifestation of the properties of a Fermi liquid.

## 1.3 The second stage of condensation of quasiparticles in helium. Bose-condensation of gyroton quasiparticles and the formation of composite bosons in helium-I.

### 1.3.1 *The mechanism of formation and structure of composite quasiparticles of the condensate fraction of helium-I.*

From the position of the "canonical" representation, the temperature of the transition of a gas of free bosons to the Bose-condensate state $T=T_{BK}$ is determined by the requirement $\mu=0$, where $\mu$ is the chemical potential of the system. The Bose particle system at a temperature $T<T_{BK}$ is separated into subsystems of particles of the above-condensate and condensate fractions, which are in equilibrium with chemical potentials equal to zero. The transition to the Bose-condensate state of a system of atomic particles when they are placed in an external field $\Phi$ is determined by the condition $\mu^{(tot)} = 0$, where $\mu^{(tot)}$ is the total specific thermodynamic potential of the system:

---

[5] From the standpoint of a quasiparticle representation, the same result can be obtained using the canonical distribution over the states of the pseudoatomic degrees of freedom of gyrotons. This approach is supposed to be presented in another work.

$$\mu^{(tot)} = \mu + \varphi = 0 \qquad (8)$$

Here $\varphi$ is the potential energy of the system in the field $\Phi$ per one particle. The requirement (8), obviously, should remain valid for a system of free fermions undergoing a transition to a Bose-condensate state.

According to the concepts introduced by us, *L*-pseudofermions (gyrotons) in the temperature range $T_0 \leq T < T_{cr}$ are assigned to helium atoms, which are in the field $\Phi$ of the interatomic interaction. The potential energy of atoms, "providing" the formation of $N_e$ gyroton quasiparticles, is determined by Eq.(1). The field $\Phi$ is an external field with respect to the system of free fermionic gyrotons. From this point of view, the total specific thermodynamic potential $\mu^{(tot)}$ of the gyroton system of *L*-pseudoelectrons with a chemical potential $\mu(N_L)_{T \geq T_0}$ equal to the Fermi energy $E_{F_L}(N_L)$ is determined by the relation (see Eq. (1, 3)):

$$\mu^{(tot)}(N_L)_{T \geq T_0} = \frac{N_L}{N_0} \cdot k_B T - \frac{N_e}{N_0} \cdot (1/2) k_B T = -\frac{N_0 - N_L}{N_0} \cdot (1/2) k_B T \qquad (9)$$

At $T=T_0$, according to Subsection 1.2.1, $N_L=N_{L0}=N_0$. So, the value of the total specific thermodynamic potential $\mu^{(tot)}_{T=T_0}(N_{L0})$ (Eq. (8)) is equal to zero at $T=T_0$. Consequently, the temperature $T_0$ corresponds to the temperature of the transition of the system of gyrotons to the Bose-condensate state.

At temperature $T_0$, all *L*-pseudoelectrons of bigyroton pseudoatoms occupy the states of an absolutely degenerate system of fermions in pairs. The energy $k_B T$ of the spin motion of *L*-pseudofermions located on the Fermi surface at the current temperature $T<T_0$ becomes less than the interatomic interaction energy $|\varphi_0(\upsilon_0)| = k_B T_0$ in absolute value. This means that the spin states of the corresponding bigyrotons cease to be independent. As a result, new (secondary) pair bonds begin to form between the bigyrotons by pairing *L*-pseudoelectrons that are part of two neighboring primary pairs. Moreover, only *Lp*-type particles participate in the formation of paired states of *L*-pseudoelectrons up to a certain characteristic temperature $T_c$ (see Figure 1). As a result, two gyrotons located in different (neighboring) primary pairs form secondary *p*-pairs. Thus, there is a pairing of two primary pairs and the formation of pair-pair states of gyrotons. In this case, the *L*-pseudoelectrons in the primary pairs also lose their mutual independence. From this point of view, we can say that secondary bonds are also established between the gyrotons of each primary pair at $T<T_0$ by *s*-pairing their *L*-pseudoelectrons. This means that the pair-pair states of gyrotons are determined by the superposition of states of *s*- and *p*-pseudoelectron pairs. From the standpoint of the principle of similarity of the structure of a liquid (see p. 4), these processes lead to the formation of an atomic-molecular quasiparticle with an atom-like *Lp*-subshell common to four gyrotons. We will denote such quasiparticles as "*bigytaton doublets*" or just "*doublets*".

Doublets are composite bosons. As a consequence, the process of separation of the system into an above-condensate fraction $N_0 - N_c = N_{nc}$ fermions and a condensate fraction of bosons formed on the basis of $N_c$ gyrotons begins at $T<T_0$. This stage of condensation is realized in the temperature range $T_c \leq T \leq T_0$ until all *p*-type *L*-pseudoelectrons form paired secondary pairs. And, thus, pass into the Bose-condensate state in the form of bigyroton doublets. The formation of a Bose-condensate fraction of bigyroton doublets causes the effect of an implicit separation of the motion of gyroatomic quasiparticles according to pseudospin and pseudoatom degrees of freedom.

Structurally, condensate doublets are Delaunay simplices in the form of tetrahedra [5,6]. Delaunay simplexes are the main structural elements of the incoherent condensate fraction of helium-I.

### 1.3.2 *Temperature of separation of a system of pair-correlated gyrotons into Bose-condensate and above-condensate fermion fractions.*

From the point of view of thermodynamics, the transition of the system to the Bose-condensate state is indistinguishable from the process of the "classical" gas-liquid transition [7]. Consequently, the process of Bose-condensation of helium-4 fermionic quasiparticles should proceed along the line of coexistence of the gaseous and liquid aggregate states of helium $P=P_{SVP}(T)$, where $P_{SVP}$ is the saturated vapor pressure.

Fig. 2 shows the dependence of the specific volume of liquid helium-4 on temperature at the pressure of saturated vapor, which is constructed on the basis of the data on the density of liquid helium recommended in [8]. The same figure shows the dependence of the specific volume $\upsilon$ of an absolutely degenerate system of particles with mass $M=M_{He}$ and spin $s=1/2$ on the Fermi temperature $T_F$. According to relations (1), (3)-(5), the temperature $T_0$ of helium at the point of the phase transition of fermionic gyrotons to the state of the Bose-condensate is equal to the Fermi temperature $T_F$ of the $N_0$ secondary pseudoelectron. From this condition, we obtain the values of the temperature $T_0$ and the corresponding specific volume $\upsilon_0$:

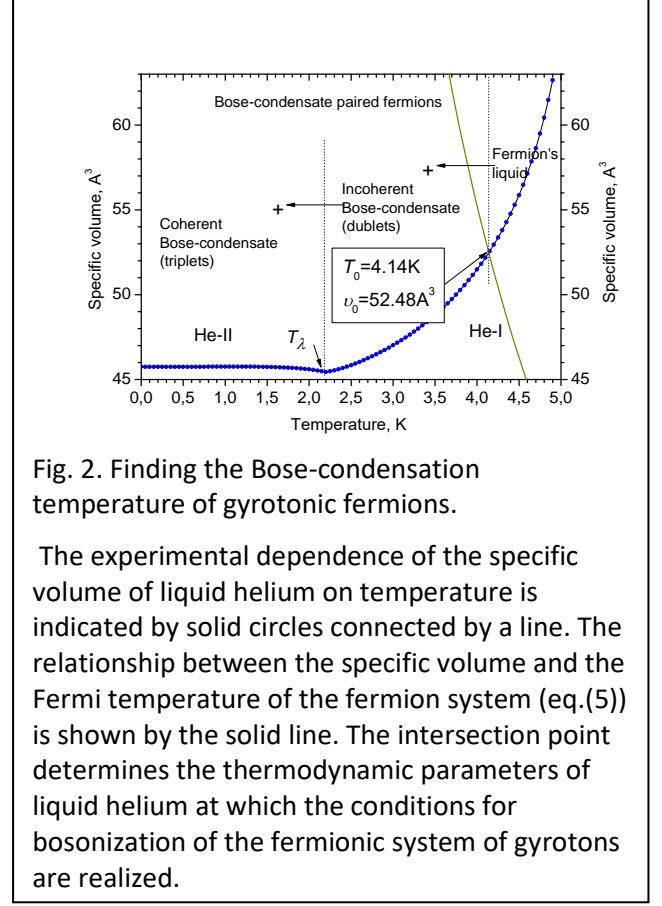

Fig. 2. Finding the Bose-condensation temperature of gyrotonic fermions.

The experimental dependence of the specific volume of liquid helium on temperature is indicated by solid circles connected by a line. The relationship between the specific volume and the Fermi temperature of the fermion system (eq.(5)) is shown by the solid line. The intersection point determines the thermodynamic parameters of liquid helium at which the conditions for bosonization of the fermionic system of gyrotons are realized.

$$T_0 = 4.14 K$$
$$\upsilon_0 = 52.48 \text{Å}^3 \quad (10)$$

The result (10) indicates that the process of Bose-condensation in helium-I and the transition of the system to a quantum-coherent state refer to different phase transitions.

According to formulas (2), (4) and (5), the temperature $T_0$ of the transition of an ensemble of gyroton quasiparticles to the Bose-condensate state is related to the critical temperature of helium-4 as $T_0 = 2^{-1/3} T_{cr}$. Substituting the value $T_0 = 4.14 K$ into this ratio, we obtain a value that almost exactly coincides with the experimental value of the critical temperature of helium-4 (see, for example, [8]):

$$T_{cr} = 2^{1/3} T_0 \approx 5.22 K \quad (11)$$

The obtained result confirms the validity of the model ideas described above about the microscopic processes occurring in helium-4 when the temperature decreases.

## 1.4 The third stage of condensation of gyroton quasiparticles in helium. The transition of helium-I into a coherent state of helium-II. Atomic-molecular liquid of three-paired gyroton quasiparticles.

### 1.4.1 *Mechanism of separation of the condensate fraction of helium-II into coherent and incoherent subsystems. Structure of composite quasiparticles of the coherent component of the condensate fraction.*

At a temperature of $T < T_c$, the bigyrotons of the above-condensate fraction carry $L$-pseudoelectrons of the $Ls$-type only. In accordance with the logic of the concepts being developed, the process of bosonization of the system should continue by pairing those of bigyroton pseudoatoms whose $L$-pseudoelectrons are on the Fermi surface at the current temperature $T$. From the point of view of the pseudoatomic bigyroton representation, the corresponding primary pairs are in a singlet state with completed pseudoatomic $Ls$ subshells. In this case, the establishment of paired bonds between two bigyrotons by the formation of hybrid $sp$ pseudomolecular orbitals of two excited $Ls$-type bigyrotons turns out to be energetically unfavorable. However, the bosonization of the fermionic states of gyrotons is possible through the mediation of previously formed doublets of the Bose-condensate fraction of helium-I.

At a temperature $T=T_c$, the proportion of gyrotons in the doublet Bose-condensate fraction is roughly 2/3 of all particles (see the results of calculations (7)). And, accordingly, the number of fermionic gyrotons of the above-condensate fraction is approximately 1/3 of all gyroton quasiparticles. This makes it possible to consider the condensate fraction of the bigyroton doublets as a kind of "matrix" in which the bigyrotons of the above-condensate fraction are "dissolved" in a ratio of 1:1 approximately. Thus, at a temperature $T=T_c$, each above-condensate bigyroton pseudoatom in the equilibrium state is surrounded by atom-like doublets of the condensate fraction. And vice versa, the closest neighbors of doublet pseudoatoms are above-condensate bigyrotons. In this case, at the temperature $T \approx T_c$, the areas of correlation interaction of condensate quasiparticles overlap throughout the entire volume. This leads to bosonization of the above-condensate $Ls$-type quasiparticles due to the formation of coherently interconnected new quasiparticle structural elements of liquid helium with decreasing temperature.

From the standpoint of the principle of AM-similarity of the structure of a liquid (see p. 4), the mechanism of bosonization of above-condensate $Ls$-type quasiparticles in helium-II is associated with the effect of $sp$ hybridization of the pseudoelectron orbitals of the condensate doublet and the orbitals of the neighboring above-condensate bigyroton and their generalization. As a result, atomic-molecular quasiparticles in the form of pair-correlated triplets of gyrotons are formed on the basis of the singlet and doublet of bigyrotons. Such quasiparticles will be called "*bigyroton triplets*" or just "*triplets*". Herewith, the openness of the pseudoelectron $L$-shell of the atom-like state of triplet quasiparticles determines the establishment of a correlation between triplets and, thereby, the formation of quasi-polymer chains on their basis[6].

The condensate fraction $N_c$ of helium-II particles separates into two subsystems due to the formation of new structural units of the liquid. The first subsystem consists of "synthesized" earlier in phase I bigyrotonic doublets, which is formed by $N_{b2}=N_c-3(N_c-N_{0p})$ gyroton quasiparticles. The second subsystem consists of coherently coupled bigyrotonic triplets "synthesized" in phase II, which are formed by $N_{b3}=3(N_c-N_{0p})$ gyrotons by combining $(N_c-N_{0p})$

---

[6] Within the framework of the developed logic, several variants of bosonization of $Ls$-quasiparticles are possible through the mediation of bigyroton doublets of the condensate fraction. However, all of them lead, in fact, to the same result of the formation of triplets of bigyroton quasiparticles. Therefore, we will not consider in detail the features of the formation of the structure of composite quasiparticles of the ordered component of the helium-II condensate fraction in this work.

*Ls*-particles with 2($N_c$–$N_{0p}$) condensate *Lp*-particles. This separation of the condensate fraction into two components causes a jump in the heat capacity of the system at a temperature $T=T_c$ (see Part 2, Subsection 2.3.3, item *d*). This means that at $T=T_c$, a second-order phase transition occurs in the system into an ordered state of singlets and doublets relative to each other within the framework of triplet structures. The degree of order $\chi$ in the system as a whole increases from 0 to 1 as the number of triplet quasiparticles increases. Thus, the temperature $T_c$ is nothing more than the temperature $T_\lambda$ of the transition of liquid helium-4 from the state "helium-I" to the coherent ordered state "helium-II".

### 1.4.2 *Temperature of transition of helium-I into the coherent state of helium-II.*

At a temperature $T=T_0$, the number of realized quasi-independent *L*-pseudoelectrons becomes equal to the number of particles of the atomic ensemble. With a further decrease in temperature, the number of free *L*-pseudoelectrons decreases as a result of their secondary pairing and condensation at the Fermi level of the *L*-band, corresponding to the pair bond energy $\varphi_0$. Consequently, unfilled levels of energy states appear inside a Fermi sphere of radius $k_0$ (Eq.(5)) at a temperature $T<T_0$. This means that the effective temperature of the fermions becomes nonzero.

The transition of $N_c=N_0-N_{nc}$ fermions to the Bose-condensate state of composite bosons excludes the corresponding helium atoms from the formation of quasi-independent pseudospin states of *L*-pseudoelectrons (due to the saturated nature of the bonds between the structural elements of atomic-molecular quasiparticles). Therefore, the realization $N_{nc}=N_L$ of above-condensate *L*-pseudoelectronic states in the temperature range $T<T_0$ is determined by the modulus of the interaction energy $|\varphi_{nc}|$ only between $N_{nc}$ above-condensate quasiparticles:

$$|\varphi_{nc}| = (N_L/N_{nc}) \cdot k_B T = k_B T \tag{12}$$

This means that the external field of individually interacting helium atoms, in which the above-condensate pseudofermions are located, is determined by the interaction only between the $N_{nc}$ atoms themselves. In this case, the condition for the equilibrium of the above-condensate and condensate subsystems is the equality to zero of the total specific thermodynamic potential of the above-condensate fraction of fermions at $T<T_0$:

$$\mu_{nc} + \varphi_{nc} = 0 \tag{13}$$

Accordingly, the chemical potential of fermions in the temperature range $T<T_0$ is equal to the modulus of the interaction energy of above-condensate helium atoms and is determined only by the thermodynamic temperature of the system (in contrast to the case $T>T_0$, see eq. (3)):

$$\mu(N_{nc})_{T \leq T_0} = E_{F_L}(N_{nc})_{T \leq T_0} = -\varphi_{nc} = \frac{N_L \cdot k_B T}{N_{nc}} = k_B T \tag{14}$$

The temperature $T_\lambda$ of the transition of the system to the coherent state corresponds to the temperature of the completion of the transition of all *Lp* particles of the system to the condensate state of doublets. To estimate the value of $T_\lambda$, we assume that in the temperature range $T<T_0$ the effective temperature of the fermions of the above-condensate fraction remains zero (as was the case in the temperature range $T_{cr}>T \geq T_0$). In this representation, the relationship between the temperature and the number of above-condensate particles has a form similar to expression (5):

$$E_{F_L}(N_{nc})_{T \leq T_0} = k_B T_{T \leq T_0} = \frac{\hbar^2 k_{nc}^2}{2M_{He}} = \frac{\hbar^2}{2M_{He}} \left( \frac{3\pi^2 N_{nc}}{V} \right)^{2/3} \tag{15}$$

Assuming the volumes in relations (15) and (5) to be the same $V=V_0=const$, and $N_{nc}(T_\lambda)=N_{0s}$, we obtain the following estimate for the temperature of the transition of the system to a coherent state:

$$T'_\lambda = T_0 \left(N_{0s}/N_0\right)^{2/3} = 2.197K \qquad (16)$$

We denoted the obtained estimate as $T'_\lambda$. The deviation of the value $T'_\lambda$ from the experimental value of the temperature of the transition of liquid helium-4 to the coherent state is about 1% ($T_\lambda=2.175\pm0.005K$, see, for example, [8]). To determine the exact value of the temperature of the phase transition of liquid helium from the "helium-I" state to the "helium-II" state, it is necessary to obtain an explicit expression for the relationship between the number of above-condensate particles and the temperature of the system. It is not possible to obtain such a relation in the entire temperature range $T \leq T_0$ in an analytical form. However, in the temperature range $T \leq T'_\lambda$, the desired relationship can be obtained with good accuracy in the explicit form[7]:

$$T = \alpha_0 T_0 \left(N_{nc}/N_0\right)^{2/3} \qquad (17a)$$

where $\alpha_0 \approx 0.989$. Substituting the values of $\alpha_0$ and $N_{0s}/N_0$ into formula (17a), we obtain the temperature $T_\lambda$, which practically exactly coincides with the experimental temperature of the transition of helium-4 into the coherent state "helium-II":

$$T_\lambda = \alpha_0 T_0 \left(\frac{N_{0s}}{N_0}\right)^{2/3} = \alpha_0 T'_\lambda = 2.174K \qquad (17b)$$

## 1.5 Results of the first part of the work

In the first part of the paper, the basic principles and ideas of microscopic physics of a macroscopic ensemble of helium atoms are formulated. On this basis, an evolutionary structural-dynamic model of liquid helium is constructed. Within the framework of this model, the conditions, atomic-molecular forms and characteristic thermodynamic parameters of the transitions of the system from the gaseous state to the liquid state, then to the Bose-condensate state and, finally, to the ordered coherent state are determined. And the relationship between these parameters has also been established.

The results of calculations of the temperatures of phase transitions of the first and second order ($T_{cr}=5.22K$ and $T_\lambda=2.174K$, respectively) indicate that the ideas on which the microscopic physics of liquid helium is built in this work adequately reflect the physical processes occurring in helium-4 in the temperature range of the existence of the liquid phase. Including its transition to a coherent state with a long-range order at the point $T_\lambda$.

The above ideas about the microscopic nature and mechanisms of interaction of $^4$He atomic particles are used in the second part of the work to calculate the heat capacity of the "helium-I" and "helium-II" phases. In this paper, we will limit ourselves to calculating the heat capacity of liquid helium only in the vicinity of the $\lambda$-point the transition of the system to a coherent state.

---

[7] In this paper, the getting of the expression (17a) is omitted to reduce the length of the article. The corresponding calculations are supposed to be given in another paper.

## Part 2. The heat capacity of liquid helium-4 near the lambda point.

> *"… A… description of the behavior of liquid helium near the lambda point is given by the empirical formula:*
>
> $$C_V \approx \begin{cases} a + b\ln|T - T_\lambda|, & T < T_\lambda \\ a' + b\ln|T - T_\lambda|, & T > T_\lambda \end{cases}$$
>
> *The explanation of this behavior is left as an exercise for reader. If successful, publish! "*
>
>
>
> ***R.P. Feynman, p.34 in "Statistical Mechanics, 1972 Westview Press"***

### 2.1 Method for calculating the heat capacity of liquid helium at a temperature below the temperature $T_0$ of the transition of gyroton quasiparticles to the Bose-condensate state

From the standpoint of the quasiparticle representation liquid helium at a temperature $T=T_0$ is a two-component system of bigyroton pseudoatoms formed by $N_{0s}$ and $N_{0p}$ quasiparticles of $s$- and $p$-types, respectively. At $T<T_0$, these particles are distributed between the fraction of above-condensate pair-correlated fermionic quasiparticles (bigyrotons) and the fraction of Bose-condensate atomic-molecular quasiparticles. At the current temperature $T$, the above-condensate and condensate fractions occupy volumes proportional to the number of particles $N_{nc}$ and $N_c$ forming these fractions, respectively,:

$$V_{nc} = \upsilon N_{nc}; \quad V_c = \upsilon N_c; \quad V = V_{nc} + V_c; \quad N_0 = N_{nc} + N_c \tag{18}$$

where $\upsilon$ is the specific volume of the liquid.

To calculate the heat capacity of helium as a function of temperature, we will find the entropy of of the above-condensate and condensate fractions $S_{nc}$ and $S_c$ respectivel). And also the entropy $S_{cnc}$ of mixing of their particles/ The total entropy of the system is:

$$S = S_{nc} + S_c + S_{cnc} \tag{19}$$

The entropies of the above-condensate and condensate fractions are determined by the processes considered in the first part of this work, which form fermion pair-correlated gyrotons and the atomic-molecular quasiparticle structure of their composite bosonic states. The contributions of the individual components of the total entropy can be represented as explicit functions of the number of above-condensate quasiparticles $N_{nc}$. The resulting heat capacity of the system can be represented as the sum of the derivatives of the individual entropy terms with respect to the number of above-condensate gyroton quasiparticles. Calculations of the heat capacity of the $N_0$-particle macroensemble will be performed per one atomic particle. Using the temperature dependence (17a) of the number of above-condensate particles $N_{nc}$ we obtain the heat capacity of liquid helium as a function of the number of above-condensate quasiparticles:

$$c = \frac{T}{N_0}\frac{dS}{dT} = \frac{T}{N_0}\sum_i\left(\frac{dS_i}{dT}\right) = \left(\frac{T}{N_0}\frac{dN_{nc}}{dT}\right)\sum_i\left(\frac{dS_i}{dN_{nc}}\right) = \frac{3}{2}\left(\frac{N_{nc}}{N_0}\right)\cdot\sum_i\left(\frac{dS_i}{dN_{nc}}\right) = \sum_i c_i \tag{20}$$

Here, the subscript $i$ at the symbols $S_i$ and $c_i$ numbers the contributions of individual structural-dynamic subsystems to the total entropy $S$ and heat capacity $c$ of liquid helium-4.

To calculate the entropies of the subsystems $S_i$, we will find the statistical weights $W_i$ of the corresponding states: $S_i = \ln W_i$. According to this definition6 the entropy $S$ and heat capacity $c$ are calculated here in units of the Boltzmann constant $k_B$.

The temperature dependence (17a) of the number above-condensate particles $N_{nc}$ was obtained in Subsection 1.4.2 under the condition $V=const$. Therefore, the heat capacity determined in (20) is the heat capacity at constant volume ($c=c_V$). The heat capacity $c_s$ of liquid helium-4 in the experiment is determined at the saturated vapor pressure (*SVP*). The heat capacity $c_s$ in the vicinity of temperature $T=T_\lambda$ is practically equal to the heat capacity $c_p$ at constant pressure [9]:

$$c_s \simeq c_P = \frac{1}{N_0}\left[\left(\frac{dE}{dT}\right)_V + P_{SVP}\left(\frac{dV}{dT}\right)_{SVP}\right] = c_V + \Delta c \qquad (21)$$

Therefore, when comparing theoretical calculations of heat capacity with experimental data, one should take into account the contribution $\Delta c$, which is determined by the volume derivative with respect to the temperature at the saturated vapor pressure:

$$\Delta c = P_{SVP}\left(\frac{dV}{dT}\right)_{SVP} = \frac{T}{N_0}\left(\frac{\partial S}{\partial V}\right)_E \left(\frac{dV}{dT}\right)_{SVP} \qquad (22)$$

Thermodynamic quantities related to the area $T>T_\lambda$ (helium-I) or $T<T_\lambda$ (helium-II) will be accompanied by a subscript "+" or "-", respectively. The temperature region in the immediate vicinity of the point $T=T_\lambda$ will be denoted as $T \to T_{\lambda+}$ for $T>T_\lambda$ or $T \to T_{\lambda-}$ for $T<T_\lambda$.

## 2.2 Heat capacity of helium-I.

### 2.2.1 *The structure of the entropy $S_+$ of helium-I at temperatures below the Bose-condensation temperature $T_0$.*

The entropy of both the above-condensate and condensate fractions of helium-I is the sum of the entropies determined by the distribution of particles over the pseudospin degrees of freedom of gyrotons, and the entropies determined by the formation of internal (pseudoatomic) degrees of freedom of composite gyroton quasiparticles.

The above-condensate fraction of pair-correlated states of gyrotons at $T_0 > T \geq T_\lambda$ is formed by quasiparticles of the *Ls*- and *Lp*- components. Therefore, the entropy $S_{nc+}$ of the above-condensate fraction of helium-I can be represented as the sum of two contributions. The first contribution $S_{nc+}^{(sp)}$ is determined by the distribution of the particles of the above-condensate fraction over the states of particles of the *Ls* and *Lp* components. The second contribution $S_{nc+}^{(b1)}$ is due to the presence of primary pair bonds between gyrotons: $S_{nc+} = S_{nc+}^{(sp)} + S_{nc+}^{(b1)}$.

In turn, the condensate doublet fraction of helium-I is one-component and is formed on the basis of only *p*-type quasiparticles. Therefore, entropy $S_{c+}$ is determined only by the processes of formation of primary and secondary pairs. Accordingly, the entropy of formation of doublet structural units of the condensate fraction can be represented as the sum of the entropy $S_{c+}^{(b1)}$ of primary and $S_{c+}^{(b2)}$ secondary pairing: $S_{c+} = S_{c+}^{(b1)} + S_{c+}^{(b2)}$. Thus, in accordance with (19), the total entropy of liquid helium-I at a temperature $T<T_0$ can be represented by the sum of contributions:

$$S_+ = S_{cnc+} + S_{nc+}^{(sp)} + S_{nc+}^{(b1)} + S_{c+}^{(b1)} + S_{c+}^{(b2)} \qquad (23)$$

### 2.2.2 *Heat capacity of helium-I at saturated vapor pressure.*

The heat capacity of helium-I at a saturated vapor pressure includes the contribution $\Delta c = \Delta c_+$ (see Eq. (22)), which is determined by the derivative of the entropy $S_+$ (Eq. (23)) with respect to the volume. Herewith, the internal (pseudoatomic) degrees of freedom of composite quasiparticles do not contribute to the pressure. And, thus, the corresponding components $S_{nc+}^{(b1)}$, $S_{c+}^{(b1)}$ and $S_{c+}^{(b2)}$ of the total entropy $S_+$ do not contribute to the term $\Delta c_+$ of the heat capacity $c_{s+}$. The entropy component $S_{cnc+}$ of total entropy $S_+$ is determined only by the distribution of the particles of the condensate fraction among the particles of the above-condensate fraction at a constant number of particles of the system $N_0$ and does not depend on the volume explicitly. Therefore, the entropy component $S_{cnc+}$ also does not contribute to $\Delta c_+$. Thus, only the $S_{nc+}^{(sp)}$ component of the entropy $S_+$ contributes to $\Delta c_+$ (see Eq. (23)). The entropy $S_{nc+}^{(sp)}$ is determined by the distribution of the particles of the above-condensate fraction over the pseudospin states of *Ls*- and *Lp*-pseudofermions in the variable volume $V_{nc} = \upsilon N_{nc}$ corresponding to this subsystem (see (18)).

The component $\Delta c_+$ of the heat capacity of helium-I (Eq. (22)) can be represented as:

$$\Delta c_+ = \frac{T}{N_0} \frac{dS_{nc+}^{(sp)}}{dN_{nc}} \frac{dN_{nc}}{dV_{nc}} \left( \frac{N_0 d\upsilon_+}{dN_{nc}} \frac{dN_{nc}}{dT} \right)_{SVP} = \left( \frac{T}{N_0} \frac{dN_{nc}}{dT} \frac{dS_{nc+}^{(sp)}}{dN_{nc}} \right) \left( \frac{N_0 d\upsilon_+}{\upsilon dN_{nc}} \right)_{SVP} = c_{nc+}^{(sp)} \delta_+ \qquad (24)$$

The first factor $c_{nc+}^{(sp)}$ on the right-hand side of Eq. (24) determines the contribution to the heat capacity associated with the process of the distribution of above-condensate particles over the states of the *Ls*- and *Lp*-type components at constant volume (see below, item *b* of Subsection 2.2.3, Eq. (30) ). The second factor $\delta_+(T) = \left( \frac{N_0 d\upsilon_+}{\upsilon dN_{nc}} \right)_{SVP}$ is determined by the decrease in the volume of helium-I as a result of Bose-condensation of fermionic *Lp*-quasiparticles.

Thus, the heat capacity $c_{s+}$ of helium-I at a saturated vapor pressure is determined by the sum of contributions:

$$c_{s+}(T) \simeq \left( \frac{T}{N_0} \frac{dN_{nc}}{dT} \right) \cdot \left( \frac{dS_{cnc}}{dN_{nc}} + \frac{dS_{nc+}^{(sp)}}{dN_{nc}} + \frac{dS_{nc+}^{(b1)}}{dN_{nc}} + \frac{dS_{c+}^{(b1)}}{dN_{nc}} + \frac{dS_{c+}^{(b2)}}{dN_{nc}} \right) + \Delta c_+ =$$
$$= c_{cnc+} + c_{nc+}^{(b1)} + c_{c+}^{(b1)} + c_{c+}^{(b2)} + c_{nc+}^{(sp)} \cdot \gamma_+(T) \qquad (25)$$

Here the coefficient $\gamma_+$ at the contribution $c_{nc+}^{(sp)}$ is equal to:

$$\gamma_+ = 1 + \delta_+ = 1 + \frac{N_0}{\upsilon} \frac{d\upsilon_+}{dN_{nc}} \qquad (26)$$

### 2.2.3 *Statistical weight, entropy and heat capacity of the processes of formation of the structure of quasiparticle subsystems of helium-I near the λ-point.*

#### a. *Mixing of above-condensate and condensate fractions of helium-I.*

To determine the entropy $S_{cnc+}$ of mixing of particles of the above-condensate and condensate fractions of helium-I in the vicinity of the $\lambda$-point, one can write the expression:

$$S_{cnc+} = \ln W_{cnc+} \approx \ln \frac{N_0!}{(N_0 - N_{nc})! N_{nc}!} \tag{27}$$

Here $W_{cnc+}$ is the statistical weight of the mixing of particles of the above-condensate and condensate helium-I fractions.

According to Eq.(20), we obtain the contribution of the mixing process of condensate and above-condensate fractions to the heat capacity of helium-I in the vicinity of the $\lambda$-point from Eq.(27), using the Stirling formula:

$$c_{cnc+} = \frac{T}{N_0} \frac{dN_{nc}}{dT} \cdot \frac{dS_{cnc+}}{dN_{nc}} = \left(\frac{3}{2} \frac{N_{nc}}{N_0}\right) \cdot \ln\left(\frac{N_0 - N_{nc}}{N_{nc}}\right) \tag{28}$$

All the following calculations of individual contributions to entropy and heat capacity are performed in a similar way.

#### b. *Distribution gyrotons of the above-condensate fraction of helium-I over the Ls - and Lp-states.*

The statistical weight $W_{nc+}^{(sp)}$, entropy $S_{nc+}^{(sp)}$, and heat capacity $c_{nc+}^{(sp)}$ of the process of distribution of $N_{nc}$ above-condensate particles over the *Ls*- and *Lp*-type states near the $\lambda$-point are determined by the relations:

$$S_{nc+}^{(sp)} = \ln W_{nc+}^{(sp)} = \ln \frac{N_{nc}!}{(N_{nc} - N_{0s})! N_{0s}!} \tag{29}$$

$$c_{nc+}^{(sp)} = \frac{T}{N_0} \frac{dN_{nc}}{dT} \cdot \frac{dS_{nc+}^{(sp)}}{dN_{nc}} = \left(\frac{3}{2} \frac{N_{nc}}{N_0}\right) \cdot \ln\left(\frac{N_{nc}}{N_{nc} - N_{0s}}\right) \tag{30}$$

As can be seen, the contribution (30) to the heat capacity of helium-I tends to infinity at $N_{nc} \rightarrow N_{0s}$ and, therefore, at $T \rightarrow T_{\lambda+}$.

#### c. *Contribution of the formation of primary pairs of gyrotons of the above-condensate fraction to the heat capacity of helium-I.*

The formation of primary pseudofermion pairs is determined by the pairing of *K*-pseudoelectrons. Herewith, the degree of degeneracy of the pseudofermion states of primary pairs of gyrotons is determined by the degree of spin degeneracy $g_s=2$ of the states of *L*-pseudoelectrons generated by induced secondary pseudospins localized on helium atoms. Almost all *L*-pseudoelectrons of the above-condensate fraction of bigyroton pseudoatoms are *s*-type quasiparticles at $T \rightarrow T_{\lambda+}$. In this case, the contribution of the internal degrees of freedom of the above-condensate fraction to the entropy and heat capacity of helium-I in the vicinity of the $\lambda$-point is found as follows:

$$S_{nc+}^{(b1)} = \ln W_{nc+}^{(b1)} = \ln \left\{ \frac{(N_{nc})!}{[(N_{nc}/2)!]^2} \right\}^{g_s} \approx 2 N_{nc} \ln 2 \tag{31}$$

$$c_{nc+}^{(b1)} = \left(\frac{T}{N_0}\frac{dN_{nc}}{dT}\right) \cdot \frac{dS_{nc+}^{(b1)}}{dN_{nc}} = \frac{3}{2}\left(\frac{N_{nc}}{N_0}\right) \cdot \ln 4 \qquad (32)$$

### d. Contribution of the internal degrees of freedom of the condensate fraction to the heat capacity of helium-I.

The formation of doublet quasiparticles of the condensate fraction is determined by the processes of formation of both primary and secondary gyroton pairs (see Section 1.3.1). The formation of primary pairs of pseudofermions occurs in the entire ensemble of $N_0$ atomic particles by pairing $K$-pseudoelectrons. Therefore, including the contribution of primary pairing to the entropy and heat capacity of the above-condensate fraction, we should also include the contribution of primary pairing to the entropy and heat capacity of the condensate helium fraction. Primary pairing includes the process of localization of secondary pseudospins on helium-I atoms, which determines the "conservation" of the fermionic properties of individual gyrotons. Thus, when calculating the entropy $S_{c+}^{(b1)}$ and heat capacity $c_{c+}^{(b1)}$, we must, as in the case of above-condensate quasiparticles, take into account the degree of spin degeneracy of primary pairs, which precedes their transition to the condensate state by secondary pairing. These contributions are determined similarly to the corresponding contributions of the above-condensate fraction at $T \to T_{\lambda+}$ (see Eq. (31,32)):

$$S_{c+}^{(b1)} = \ln W_{c+}^{(b1)} = \ln\left\{\frac{(N_c)!}{\left[(N_c/2)!\right]^2}\right\}^{g_s} \approx 2N_c \ln 2 \qquad (33)$$

$$c_{c+}^{(b1)} = \left(\frac{T}{N_0}\frac{dN_{nc}}{dT}\right) \cdot \left(\frac{d(S_{c+}^{b1})}{dN_{nc}}\right) = -\frac{3}{2}\left(\frac{N_{nc}}{N_0}\right) \cdot \ln 4 \qquad (34)$$

Here $g_s = 2$ is the spin degree of degeneracy of the states of the primary pairs, which is determined by unpaired secondary pseudospins $s=1/2$ localized on atoms.

According to Subsection 1.3.1, the actual transition of gyrotons to the Bose-condensate state is associated with the process of secondary pairing of pair-correlated quasiparticles. The physical content of this process is due to the pairing of primary pairs through the formation of common $Lp$-orbitals of $L$-pseudoelectrons of the doublet pseudomolecular quasiparticle. As a result, the pseudospin states of four gyrotons cease to be independent.

The states of the bigyroton doublets are determined by the superposition of the states of $s$- and $p$-pseudoelectron pairs (see Subsection 1.3.1). From this point of view, the effect of the formation of doublets can be conventionally represented as the combined result of two processes. First, new bonds are formed by $p$-pairing of $Lp$-pseudoelectrons that are part of two neighboring primary pairs. The degree of degeneracy of $p$-pair states is $g_1=3$. And, secondly, new bonds are also formed between the gyroatoms of the primary pair by $s$-pairing of their $Lp$-pseudoelectrons. Herewith, the degree of degeneracy of the pseudospin state of the primary pair $g_s=2$ decreases to the value of the degree of degeneracy of the $s$-pair state of the $L$-pseudoelectrons $g_0=1$. Taking these circumstances into account, the resulting degree of degeneracy of the pair states of gyrotons formed in the process of secondary pairing will be equal to $g_a=g_1+g_0-g_s=2$. In this case, to determine the entropy $S_{c+}^{(b2)}$ and heat capacity $c_{c+}^{(b2)}$ of the process of formation of doublet states of condensate bigyrotons, we obtain the expressions:

$$S_{c+}^{(b2)} = \ln W_{c+}^{(b2)} = \ln\left\{\frac{(N_c)!}{\left[(N_c/2)!\right]^2}\right\}^{g_a} \approx 2N_c \ln 2 \tag{35}$$

$$c_{c+}^{(b2)} = \left(\frac{T}{N_0}\frac{dN_{nc}}{dT}\right)\cdot\left(\frac{d\left(S_{c+}^{b2}\right)}{dN_{nc}}\right) = -\frac{3}{2}\left(\frac{N_{nc}}{N_0}\right)\cdot\ln 4 \tag{36}$$

### 2.2.4 *Temperature dependence of the heat capacity of helium-I near the λ-point.*

According to the results of Subsections 2.2.2 and 2.2.3, the heat capacity of helium-I at a saturated vapor pressure near the temperature $T_{\lambda+}$ is determined by the equation:

$$c_{T>T_\lambda} = c_{\tau+} = \frac{3}{2}\frac{N_{nc}}{N_0}\cdot\left[\gamma_+\cdot\ln\left(\frac{N_{nc}}{N_{nc}-N_{0s}}\right) + \ln\left(\frac{N_0-N_{nc}}{N_{nc}}\right) - \ln 4\right] \tag{37}$$

Using relation (17a), we obtain the temperature dependence of the heat capacity of helium-I in the region of the λ-transition temperature, which can be represented as a linear function of the logarithm of the variable $\tau_+$:

$$c_{\tau+} = a_+ + b_+\ln(\tau_+), \tag{38}$$

$$\tau_+ = \left|(T-T_\lambda)/T_\lambda\right| = (T/T_\lambda - 1); \tag{39}$$

the coefficients $a_+$ and $b_+$ of Eq. (38) are equal, respectively:

$$a_+ = \frac{3}{2}\frac{N_{0s}}{N_0}\cdot\left[\ln\left(\frac{1}{4}\cdot\frac{N_{0p}}{N_{0s}}\right) - \gamma_+(T_\lambda)\cdot\ln\frac{3}{2}\right], \tag{40a}$$

$$b_+(T_\lambda) = -\gamma_+(T_\lambda)\cdot\frac{3}{2}\cdot\frac{N_{0s}}{N_0}; \tag{40b}$$

the values of $N_{0s}/N_0$ and $N_{0p}/N_{0s}$ are determined by relations (7).

Thus, on the basis of the principles of microscopic physics of the macroensemble of helium-4 atoms formulated in the first part of the work, a theoretical dependence of the heat capacity on temperature is obtained, which coincides in form with the empirical dependence describing the corresponding experimental data of helium-I (see the epigraph on page 16).

### 2.3 Heat capacity of helium-II.

#### 2.3.1 *The structure of the entropy $S_-$ of helium-II.*

The general scheme for calculating the heat capacity of helium-II is the same as for helium-I. The specificity of calculating the entropy of liquid helium-II is caused by the transition of the system to a state with long-range order. This state of helium-II is due to the formation of a uniform space-time structure of the internal vortex field for the entire system, which determines the interdepended distribution of gyroton quasiparticles in the states of *s*- and *p*- types.

The above-condensate and condensate fractions at the λ-point consist only of *Ls*- and *Lp*-type particles, respectively: $N_{nc}(T_\lambda)=N_{0s}$; $N_c(T_\lambda)=N_{0p}$. The degree of ordering of the system at this point is characterized by the order parameter $\chi=0$. The entropy of mixing of particles of the condensate and above-condensate fractions $S_{cnc}(T_\lambda) = S_{cnc}^{(\lambda)}$ in this case corresponds to the entropy $S^{(sp)}(T_\lambda)$ of the distribution of all above-condensate $N_{0s}$ particles $Ls$- type among all condensate $N_{0p}$ particles of the $Lp$-type.

With decreasing temperature, an increase in the number of condensate particles in the helium-II occurs due to $N_{sc}=(N_{0s}-N_{nc})=(N_c-N_{0p})$ pair-correlated $Ls$-type quasiparticles, which transform into a condensate state with the participation of $Lp$-particles of condensate doublets. As a result, an ordered subsystem $N_{b3}=3(N_c-N_{0p})$ of gyroton quasiparticles is formed in the condensate fraction of helium-II in the form of bigyrotonic triplets (see Subsection 1.4.1).

At a temperature $T<T_\lambda$, the bosonization of $Ls$-type $N_{sc}(T)$ quasiparticles and the formation of an ordered subsystem of the condensate fraction on their basis excludes these $Ls$-particles from the process of mixing of the $N_{nc}(T_\lambda)$ and $N_c(T_\lambda)$ particles of the above-condensate and condensate fractions, respectively, which took place at $T=T_\lambda$ and $\chi=0$. Therefore, to calculate the entropy $S_{cnc-}(T)$ of mixing of condensate and above-condensate fractions of helium-II with a nonzero parameter of order ($\chi \neq 0$), it is necessary to subtract the entropy $S_{c-}^{(sp)}(T)$ of mixing of $N_{sc}(T)=N_{0s}-N_{nc}(T)$ and $N_{0p}$ the condensate particles $Ls$- and $Lp$- types, respectively, from the entropy $S_{cnc}(T_\lambda)=S^{(sp)}(T_\lambda)$, which corresponds to mixing of all $Ls$- among all $Lp$- particles in the state of a system with a zero order parameter ($\chi=0$):

$$S_{cnc-}(T) = S_{cnc}(T_\lambda) - S_{c-}^{(sp)}(T) = S_{cnc}^{(\lambda)} - S_{c-}^{(sp)} \qquad (41)$$

Thus, the general structure of the entropy of helium-II, in accordance with the above considerations and according to (19, 41), can be written in the form:

$$S_- = S_{cnc}^{(\lambda)} - S_{c-}^{(sp)} + S_{nc-} + S_{c-}, \qquad (42a)$$

Below the $\lambda$-transition temperature, the above-condensate fraction of the paired states of gyrotons includes only particles of the $Ls$-component. Therefore, the contribution $S_{nc-}$ of the above-condensate fraction to the entropy of the system is determined only by the formation of internal degrees of freedom of the bigyroton pseudoatoms: $S_{nc-} = S_{nc-}^{(b1)}$. This process of establishing primary pair states occurs between all $N_0$ particles of the system. Hence, the primary pairing also contributes $S_{c-}^{(b1)}$ to the total entropy $S_{c-}$ of the condensate fraction of helium-II. In addition, the entropy condensate fraction $S_{c-}$ includes contributions $S_{c-}^{(b2)}$ and $S_{c-}^{(b3)}$, which are determined by the formation of the internal degrees of freedom of subsystems of doublet and triplet quasiparticles respectively (see Subsection 1.4.1).

The unordered subsystem of doublets of the condensate fraction of helium-II consists of $N_{b2}=N_c-3(N_c-N_{0p})$ gyroton quasiparticles of the $Lp$- component and, accordingly, does not contribute to the total entropy due to the distribution of particles over the states of different components. An ordered subsystem of triplets of the condensate fraction is formed on the basis of quasiparticles of $Ls$- and $Lp$-components. However, due to its ordered state, this subsystem also does not contribute to the entropy $S_{c-}$ of the condensate fraction of helium-II, associated with the distribution of $Ls$- and $Lp$-type particles relative to each other. The actual contribution of the ordering process to the entropy of the system is included into the entropy $S_{cnc-}(T)$ of mixing the particles of the condensate and above-condensate fractions in the form of the syntropy component $-S_{c-}^{(sp)}$ (Eq. (41)).

The values of the entropies $S_{c-}^{(b2)}$ and $S_{c-}^{(b3)}$ of the two subsystems of the helium-II condensate fraction are finite, and their contributions to the entropy of the condensate fraction is directly proportional to the number of gyroton quasiparticles forming these subsystems. The number of particles $N_{b3}=3(N_c-N_{0p})$ forming pseudomolecular triplets tends to zero at $T\to T_\lambda$. And, accordingly, the number of particles $N_{b2}=N_c-3(N_c-N_{0p})$ forming doublets tends to $N_c$. Therefore, in practice, near the $\lambda$-point, the entropy of the internal degrees of freedom of the condensate fraction of helium-II is determined only by the sum of terms $S_{c-}^{(b1)}$ and $S_{c-}^{(b2)}$. Consequently, in near the temperature of the $\lambda$-transition, the entropy of helium-II can be calculated as the sum of contributions:

$$S_- \simeq S_{cnc}^{(\lambda)} - S_{c-}^{(sp)} + S_{nc-}^{(b1)} + S_{c-}^{(b1)} + S_{c-}^{(b2)} \qquad (42b)$$

### 2.3.2 Heat capacity of helium-II at saturated vapor pressure.

Only one $-S_{c-}^{(sp)}$ component of the entropy $S_-$ (see Eq. (42b)) contributes to the $\Delta c = \Delta c_-$ term of the heat capacity $c_s$ (see Eq. (21, 22)) of helium - II for the same reasons as for helium-I (see Subsection 2.2.2). According to Subsection 2.3.1, this component reflects the effect of ordering of particles of the condensate fraction over $s$- and $p$- states in a variable volume $V_c = \upsilon N_c$ corresponding to a given subsystem (see (18)).

Thus, taking into account the above and carrying out transformations in Eq. (22) similar to those in the derivation of Eq. (24) in Subsection 2.2.2, we obtain an expression for the component $\Delta c_-$ of the atomic heat capacity $c_s$ of helium-II in the form:

$$\Delta c_- = P_{SVP}\left(\frac{dV_-}{N_0 dT}\right)_{SVP} = \left(\frac{T}{N_0}\frac{dN_{nc}}{dT}\frac{d\left(-S_{c-}^{(sp)}\right)}{dN_{nc}}\right)\left(\frac{-N_0 d\upsilon_-}{\upsilon dN_{nc}}\right)_{SVP} = c_{c-}^{(sp)} \cdot \delta_- \qquad (43)$$

Here, the first factor $c_{c-}^{(sp)}$ on the right-hand side of Eq. (43) determines the contribution to the heat capacity at constant volume associated with the process of ordering the particles of the $Ls$-component relative to the particles of the $Lp$- component (see below, item $b$ of Subsection 2.3.3, Eq. (48)). The second factor on the right-hand side of Eq. (43) $\delta_-(T) = \left(\frac{-N_0 d\upsilon_-}{\upsilon dN_{nc}}\right)_{SVP}$ is determined by an increase in the specific volume of helium-II in the process of Bose-condensation of $Ls$-fermionic quasiparticles. Thus, according to the formulas (19-23) and (41, 42b, 43), the total heat capacity of helium-II at the saturated vapor pressure in the area of the $\lambda$-transition point is determined by the sum of contributions:

$$c_{s-}(T) \simeq \left(\frac{T}{N_0}\frac{dN_{nc}}{dT}\right) \cdot \left(\frac{d\left(-S_{c-}^{(sp)}\right)}{dN_{nc}} + \frac{dS_{nc-}^{(b1)}}{dN_{nc}} + \frac{dS_{c-}^{(b1)}}{dN_{nc}} + \frac{dS_{c-}^{(b2)}}{dN_{nc}}\right) + \Delta c_- =$$
$$= c_{nc-}^{(b1)} + c_{c-}^{(b1)} + c_{c-}^{(b2)} + \gamma_-(T) \cdot c_{c-}^{(sp)} \qquad (44)$$

$$\gamma(T<T_\lambda) = \gamma_- = (1+\delta_-) = 1 - \frac{N_0}{\upsilon}\frac{d\upsilon_-}{dN_{nc}} \qquad (45)$$

### 2.3.3 Statistical weight, entropy and heat capacity of the processes of formation of the structure of quasiparticle subsystems of helium-II near the $\lambda$-point.

#### a. Mixing of the above-condensate and condensate fractions at the $\lambda$-point.

According to Subsection 2.3.1, the entropy of mixing the condensate and above-condensate fractions of helium-II $S_{cnc-}(T)$ with a non-zero parameter of order $\chi$ is determined by the difference of two terms: $S_{cnc-}(T) = S_{cnc}^{(\lambda)} - S_{c-}^{(sp)}$, (see Eq. (41)). The first component $S_{cnc}^{(\lambda)}$ of the mixing entropy $S_{cnc-}(T)$ numerically coincides with the entropy $S^{(sp)}(T_\lambda)$ (see Subsection 2.3.1):

$$S_{cnc}^{(\lambda)} = \ln W_{cnc}^{(\lambda)} = \ln\left(\frac{N_0!}{N_{0s}!N_{0p}!}\right) \quad (46)$$

This component of the mixing entropy $S_{cnc-}(T)$ does not depend on temperature and does not make a contribution to the heat capacity of helium II. In accordance with the considerations set out in Subsection 2.3.1, the second term $S_{c-}^{(sp)}$ is excluded from the first contribution due to the transition of $N_{sc}(T)=N_{0s}-N_{nc}(T)$ above-condensate $Ls$-particles to the ordered state relative to the $Lp$-particles. Thus, the temperature dependence of the entropy of mixing of the condensate and above-condensate fractions at $T<T_\lambda$ is determined by the process of formation of an ordered subsystem of the condensate fraction of helium-II. The syntropy $-S_{c-}^{(sp)}$ of this process and the corresponding contribution to the heat capacity of helium-II are given in the next item of the subsection.

### b. Contribution of the ordering process of gyroton quasiparticles of a two-component system of the condensate fraction to the heat capacity of helium-II.

The statistical weight $W_{c-}^{(sp)}$ and entropy $S_{c-}^{(sp)} = \ln W_{c-}^{(sp)}$ of the distribution of $N_{sc}(T)=N_{0s}-N_{nc}(T)$ quasiparticles of $Ls$-type among $N_{0p}$ quasiparticles of $Lp$-type of the condensate fraction are determined by the ratio:

$$S_{c-}^{(sp)} = \ln\frac{N_c!}{N_{0p}!(N_c - N_{0p})!} \quad (47)$$

Thus, according to (41), the heat capacity of the process of mixing the above-condensate and condensate fractions of helium II is determined by the derivative with respect to temperature of the syntropy of the ordering of the two-component system $-S_{c-}^{(sp)}$ at $T<T_\lambda$:

$$c_{c-}^{(sp)} = \left(\frac{T}{N_0}\frac{dN_{nc}}{dT}\right)\cdot\frac{d(-S_{c-}^{(sp)})}{dN_{nc}} = \left(\frac{3}{2}\frac{N_{nc}}{N_0}\right)\ln\left(\frac{N_c}{N_c - N_{0p}}\right) \quad (48)$$

The contribution $c_{c-}^{(sp)}$ to the heat capacity of helium-II tends to infinity at $N_{nc}\rightarrow N_{0s}$ and, therefore, at $T\rightarrow T_{\lambda-}$, (as in the case of the heat capacity $c_{nc+}^{(sp)}$ of helium-I at $T\rightarrow T_{\lambda+}$, see Eq. (30)).

### c. Contribution of the formation of primary pairs of gyrotons of the above-condensate fraction to the heat capacity of helium-II.

The expressions for determining the statistical weight $W_{nc-}^{(b1)}$, entropy $S_{nc-}^{(b1)} = \ln W_{nc-}^{(b1)}$, and heat capacity $c_{nc-}^{(b1)}$ of the process of formation of primary pairs of fermionic gyrotons of the above-condensate fraction in the vicinity of the temperature $T_\lambda$ for helium-II coincide with the expressions for helium-I (Eq. (31) and (32)):

$$S_{nc-}^{(b1)} = \ln W_{nc-}^{(b1)} = \ln\left\{\frac{N_{nc}!}{[(N_{nc}/2)!]^2}\right\} = N_{nc}\ln 4 \quad (49)$$

$$c_{nc-}^{(b1)} = \left(T\frac{d(N_{nc}/N_0)}{dT}\right)\cdot\frac{dS_{nc-}^{(b1)}}{dN_{nc}} = \left(\frac{3}{2}\frac{N_{nc}}{N_0}\right)\ln 4 \quad (50)$$

### d. Heat capacity of internal degrees of freedom of gyrotons of the condensate fraction of helium-II.

In helium-II, as in helium-I, the contribution of the internal degrees of freedom to the entropy of the condensate fraction is made by the processes of formation of both primary and secondary pairs.

The process of formation of primary fermion pairs is not associated with the transition of helium into a coherent state. As a consequence, primary pairing makes contributions to the entropy and heat capacity of the helium-II similar to the contributions to the entropy and heat capacity of the helium-I condensate fraction (see (33, 34)):

$$S_{c-}^{(b1)} = \ln W_{c-}^{(b1)} = \ln \left\{ \frac{(N_c)!}{\left[(N_c/2)!\right]^2} \right\}^2 \simeq N_c \cdot \ln 4 \qquad (51)$$

$$c_{c-}^{(b1)} = \left( T \frac{d(N_{nc}/N_0)}{dT} \right) \cdot \frac{dS_{c-}^{(b1)}}{dN_{nc}} = -\left( \frac{3}{2} \frac{N_{nc}}{N_0} \right) \cdot \ln 4 \qquad (52)$$

According to the concepts presented in Subsection 1.4.1, the transition of $Ls$ particles to the Bose-condensate state is due to the effect of their secondary pairing mediated by particles of the doublet condensate fraction. This leads to the separation of $N_c$ particles of the condensate doublet fraction of helium-II into two components – an ordered subsystem of newly formed triplet quasiparticles and unordered subsystem of previously formed doublet quasiparticles.

According to Subsection 2.3.1, the contribution of the internal degrees of freedom of the triplet subsystem to the entropy and heat capacity of the helium-II condensate fraction in the temperature range $T \to T_{\lambda-}$ can be neglected. In turn, the contribution $N_{b2}=N_c-3(N_c-N_{0p})$ of the particles of the doublet subsystem to the entropy of the helium II condensate fraction due to the secondary pairing effect is calculated similarly to the corresponding contribution to the helium-I condensate fraction entropy (see Subsection 2.2.3, item $d$, equation (35)):

$$S_{c-}^{(b2)} = \ln W_{c-}^{(b2)} = \ln \left\{ \frac{(3N_{0p} - 2N_c)!}{\left\{\left[(3N_{0p} - 2N_c)/2\right]!\right\}^2} \right\}^{g_a} \approx (3N_{0p} - 2N_c) \ln 4 \qquad (53)$$

The number of particles of the helium-II condensate fraction at $T \to T_{\lambda-}$ is almost completely determined by the number of particles of the doublet subsystem ($N_{b2} \to N_c$). Therefore, the contribution of the doublet subsystem of the condensate fraction to the heat capacity of helium-II in the temperature range $T_{\lambda-}$ is equal to:

$$c_{c-}^{(b2)} = \left( \frac{T}{N_0} \frac{dN_{nc}}{dT} \right) \cdot \frac{dS_{c-}^{(b2)}}{dN_{nc}} = \left( \frac{3}{2} \frac{N_{nc}}{N_0} \right) \cdot 2 \ln 4 \qquad (54)$$

Comparison of the contributions made by the fraction of doublet quasiparticles to the heat capacity of helium-I $c_{c+}^{(b2)}$ and helium-II $c_{c-}^{(b2)}$ (Eqs. (36) and (54), respectively) shows that the effect of separation of the condensate fraction of helium into an ordered and disordered subsystems of bigyrotons at the $\lambda$-point leads to the finite jump in heat capacity, equal, according to formulas (36, 54) and (7):

$$\delta c_\lambda = c_{c-}^{(b2)} - c_{c+}^{(b2)} = \frac{3}{2}\left(\frac{N_{0s}}{N_0}\right)\cdot 3\ln 4 = 2.411 \tag{55}$$

### 2.3.4 *Temperature dependence of the heat capacity of helium-II near the λ-point.*

According to the results of Subsection 2.3.3 and in accordance with (44, 45), the heat capacity of helium II at a saturated vapor pressure near the temperature $T_{\lambda-}$ is determined as follows:

$$c_{T<T_\lambda} = c_{\tau-} = \frac{3}{2}\frac{N_{nc}}{N_0}\left[\gamma_-(T)\cdot\ln\left(\frac{N_c}{N_c - N_{0p}}\right) + 2\ln 4\right] \tag{56}$$

Expression (56) for the temperature dependence of the heat capacity of helium-II in the vicinity of the temperature $T_\lambda$ is reduced to a form similar to (38):

$$c_{\tau-} = a_- + b_-\ln(\tau_-), \tag{57}$$

where the variable $\tau_-$ of equation (57) is determined by the expression:

$$\tau_- = |(T - T_\lambda)/T_\lambda| = (1 - T/T_\lambda). \tag{58}$$

Coefficients $a_-$ and $b_-$ of equation (57) are equal, respectively:

$$a_-(T_\lambda) = \frac{3}{2}\frac{N_{0s}}{N_0}\cdot\left[\ln\left(16\cdot\frac{N_{0p}}{N_{0s}}\right) - \gamma_-(T_\lambda)\cdot\ln\frac{3}{2}\right], \tag{59a}$$

$$b_-(T_\lambda) = -\gamma_-(T_\lambda)\cdot\frac{3}{2}\frac{N_{0s}}{N_0}; \tag{59b}$$

### 2.4 Comparison of theoretical calculations of the heat capacity of liquid helium-4 with experimental data in the vicinity of the point of the λ-transition of the system into a coherent state.

For a numerical comparison of the theoretical and experimental dependences of the heat capacity on temperature in the vicinity of the λ-point, it is necessary to determine the values of the coefficients $\gamma(T_{\lambda+}) = \gamma_+^{(\lambda)}$ and $\gamma(T_{\lambda-}) = \gamma_-^{(\lambda)}$ (Eq. (26, 45) ). For this purpose in this paper we used experimental data on the temperature dependence of the density of liquid helium at a saturated vapor pressure (see work [8]). The corresponding estimate of the coefficient $\delta(T_{\lambda+}) = \delta_+^{(\lambda)} = (1 - \gamma_+^{(\lambda)})$ when moving to the temperature of the λ-transition from the region $T > T_\lambda$ gives an approximate value equal to:

$$\delta_+ \approx 0.07 \pm 0.02 \tag{60a}$$

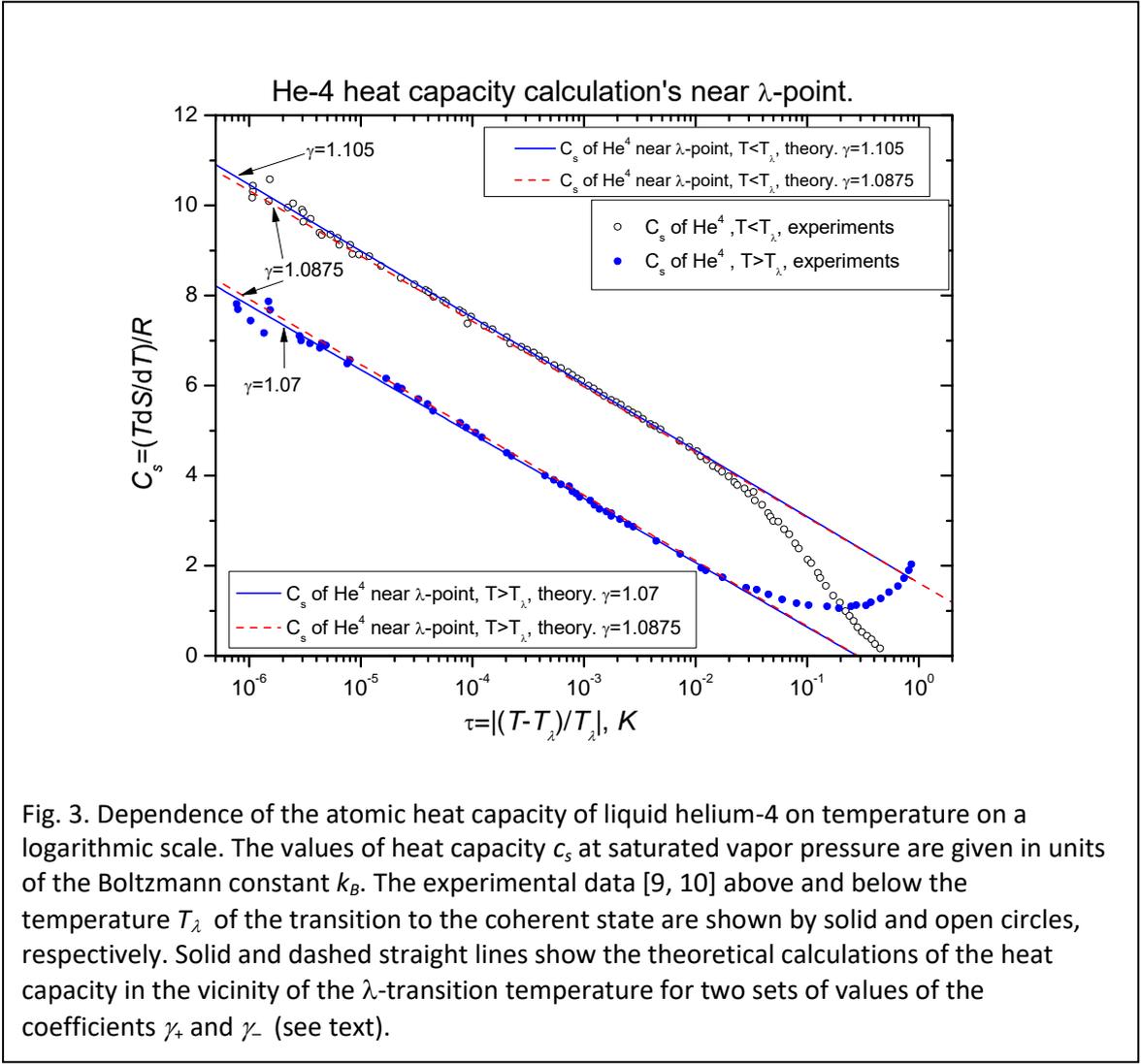

Fig. 3. Dependence of the atomic heat capacity of liquid helium-4 on temperature on a logarithmic scale. The values of heat capacity $c_s$ at saturated vapor pressure are given in units of the Boltzmann constant $k_B$. The experimental data [9, 10] above and below the temperature $T_\lambda$ of the transition to the coherent state are shown by solid and open circles, respectively. Solid and dashed straight lines show the theoretical calculations of the heat capacity in the vicinity of the $\lambda$-transition temperature for two sets of values of the coefficients $\gamma_+$ and $\gamma_-$ (see text).

The estimate of the coefficient $\delta(T_{\lambda-}) = \delta_-^{(\lambda)} = \left(1 - \gamma_-^{(\lambda)}\right)$ when moving to the temperature of the $\lambda$-transition from the region $T<T_\lambda$ gives the value:

$$\delta_- \approx 0.105 \pm 0.02 \tag{60b}$$

Substituting the estimated values $\gamma_{+/-}^{(\lambda)} = 1 + \delta_{+/-}^{(\lambda)}$ and the calculated values of the relative fractions of gyroton quasiparticles $N_{0s}/N_0$ and $N_{0p}/N_0$ of the $s$- and $p$-type respectively (see results (7), Subsection 1.2.2), into the expressions for the coefficients $a_+, b_+$ and $a_-, b_-$ (formulas (40a,b) and (59a,b)), we obtain the following relations for the temperature dependence of the atomic heat capacity of helium-I and helium-II in the vicinity of the $\lambda$-transition point (in units of $k_B$):

$$c_{\tau+} = a_+ + b_+ \ln(\tau_+) = -0.7873 - 0.6203 \cdot \ln(\tau_+) \tag{61}$$

$$c_{\tau-} = a_- + b_- \ln(\tau_-) = 1.6155 - 0.6406 \cdot \ln(\tau_-) \tag{62}$$

In fig. 3 shows the results of precision measurements of the heat capacity of liquid helium at a saturated vapor pressure (see [9, 10]). The same figure shows theoretical calculations of the heat capacity of helium-4 in the $\lambda$-transition region (Eq. (61) and (62) for $T \to T_{\lambda+}$ and $T \to T_{\lambda-}$, respectively). As can be seen, the results of heat capacity in the temperature range $|T - T_\lambda| < 10^{-2} K$ are in very good agreement with the experimental data.

As follows from (61) and (62), the coefficients $b_+$ and $b_-$ (Eq. (40$b$) and (59$b$)) at the logarithm $\tau=|T/T_\lambda-1|$ (equations (38) and (57)) differ by about 3%. The accuracy of the experimental data used to determine the coefficients $\delta_+^{(\lambda)}$ and $\delta_-^{(\lambda)}$ does not allow us to unambiguously state that the coefficients $b_+$ and $b_-$ are different[8]. For this reason, in fig. 3 also shows the results of calculating the heat capacity with the equal values of the coefficients $\gamma=1+\delta$ for the case of $T \to T_{\lambda+}$ and $T \to T_{\lambda-}$:

$$\gamma_{+/-}^{(\lambda)} = 1+\delta_{+/-}^{(\lambda)} = 1+\left(\delta_+^{(\lambda)}+\delta_-^{(\lambda)}\right)/2 = 1.0875 \tag{63}$$

This leads to the equality of the coefficients $b_+$ and $b_-$ of equations (38) and (57) respectively. In this case, the equations of temperature dependence of heat capacity take the form:

$$\begin{aligned} c_{\tau+} &= a'_+ + b'_+ \ln(\tau_+) = -0.7914 - 0.6305 \cdot \ln(\tau_+) \\ c_{\tau-} &= a'_- + b'_- \ln(\tau_-) = 1.6196 - 0.6305 \cdot \ln(\tau_-) \end{aligned} \tag{64}$$

As can be seen from fig. 3, the results of theoretical calculations of the heat capacity with the coefficients $a'_+, b'_+$ and $a'_-, b'_-$ of equations (64) change insignificantly in comparison with previous calculations with coefficients $a_+, b_+$ and $a_-, b_-$ of equations (61) and (62). And thus, the agreement between the calculated values of the specific heat and the experimental data for the case with equal coefficients $b'_+ = b'_-$ at the logarithm of the reduced temperature $\tau=(|1-T/T_\lambda|)$ (and, accordingly, with the same values of $\gamma_+^{(\lambda)} = \gamma_-^{(\lambda)} = \gamma_{+/-}^{(\lambda)} = 1+\delta_{+/-}^{(\lambda)}$) remains very good[9]

### 2.5 Results of the second part of the work

For the first time, it was possible to theoretically obtain the asymptotics of the heat capacity $c$ of the real three-dimensional system in the form $c=a+b*ln(|T-T_\lambda|)$ in the vicinity of the temperature $T_\lambda$ of a second-order phase transition. Herewith, a very good quantitative agreement between the calculation and experiment was obtained. Using liquid helium-4 as an example, the physical consistency of the ideas presented in this work are demonstrated for understanding the microscopic nature of cooperative quantum phenomena.

### Summary

This article develops a microscopic approach to describing the behavior of liquid helium-4 based on the concept of gyroatomic quasiparticles with pseudoatomic and pseudospin degrees of freedom. According to the concepts developed in this work, liquid helium is a material, the structural elements of which are formed like an electron shells of atoms and /or molecules.

The physical principles and mechanisms of the establishment and evolution of interparticle correlations of helium-4 atoms in the process of decreasing temperature formulated in this work can become, it seems, a conceptual basis for understanding the microscopic physics of known cooperative quantum effects and describing their properties from a unified point of view.

---

[8] At the same time, we can confidently assume that the sought values of both coefficients lie within the errors indicated in the results (60$a,b$).

[9] The question of the detailed course of the temperature dependence of the logarithmic singularity requires a theoretical calculation of the values of the parameters $\gamma_+$ and $\gamma_-$ of equations (61) and (62). Such a calculation, in principle, can be performed from the standpoint of the concepts of the microscopic physics of helium presented in this work. This task is the subject of a separate work.

# Основные идеи микроскопической физики жидкого гелия-4 и λ-переход в когерентное состояние.

С.И. Морозов[*,1]

*ГНЦ РФ Физико-энергетический институт им. А.И. Лейпунского,*
[*]*ведущий научный сотрудник в отставке*

**Аннотация**

В работе сформулированы основные положения микроскопической физики макроскопического ансамбля взаимодействующих атомов гелия-4. Вводится представление о вихревом поле, которое обусловлено движением частиц электронной и ядерной подсистем атомов макросистемы относительно друг друга. Полагается, что "погруженные" во внутреннее вихревое поле атомы гелия-4 приобретают свойства фермионных частиц благодаря порождаемым вихрем атомным псевдоспинам $s=1/2$. На этой базе построена эвристическая эволюционная модель гелия-4, как ансамбля атомоподобных квазичастиц с фермионными свойствами. В процессе снижения температуры происходит образование связей между псевдофермионными квазичастицами путем спаривания их псевдоспинов. Формирование на этой основе иерархии композитных квазичастиц приводит к переходу атомов гелия-4 из газовой фазы в жидкое состояние, затем в состояние некогерентного бозе-конденсата и, наконец, в состояние с дальним порядком атомоподобных квазичастиц. Определены структурные формы композитных квазичастиц в различных фазах, механизмы их образования и температуры указанных фазовых переходов ($T_{cr}=5.22K$, $T_0=4.14K$ и $T_\lambda=2.175K$ соответственно). Изложенные в данной работе идеи и расчеты легли в основу решения задачи о температурной зависимости теплоемкости жидкого гелия-4, как в фазе гелий-I, так и фазе гелий-II. Полученное решение имеет логарифмическую зависимость теплоемкости от приведенной температуры ($|1-T/T_\lambda|$) в области точки фазового перехода гелия-4 из неупорядоченного в упорядоченное состояние. Выполненные численные расчеты теплоемкости в окрестности температуры $T_\lambda$ находятся в очень хорошем согласии с экспериментальными данными.

---

[1] *e-mail: starshoy.frost@gmail.com*

# Оглавление

**Введение**

В 1932г. В. Х. Кеезом с сотрудниками обнаружили аномальный рост теплоемкости жидкого гелия-4 при подходе к температуре $T_\lambda \approx 2.175 K$ [1, 2]. Эта температура была названа им λ-точкой, состояния жидкого гелия выше и ниже температуры $T_\lambda$ были названы фазами "гелий-I" и "гелий-II" соответственно, а собственно фазовый переход – λ-переходом. Дальнейшие исследования обнаружили, что в фазе II гелий демонстрирует уникальные свойства когерентной квантовой системы, которые проявляются, в частности, в эффекте бездиссипативного течения жидкости по капиллярам [3, 4]. Понимание природы явления сверхтекучести невозможно без уяснения физики жидкого гелия на микроскопическом уровне. Решение этой задачи представляет принципиальный интерес для физики всех известных кооперативных квантовых явлений.

Проблема коллективного поведения атомов гелия-4 в жидкой фазе в данной работе решается путем представления данной системы, как ансамбля квазичастиц со свойствами фермионов. Такое представление означает, что взаимодействие между атомами может быть выражено через эффект приобретения атомами гелия-4 псевдоспиновых степеней свободы с полуцелым значением собственного механического момента. Данный эффект, в свою очередь, приводит к "превращению" взаимодействующих бозонных атомов в свободные фермионные квазичастицы.

Самоорганизация фермионных квазичастиц при понижении температуры обуславливает бозонизацию системы путем синтеза новых структурных элементов жидкости в форме атомоподобных композитных квазичастиц с их последовательным усложнением и, в конечном итоге, переходом в когерентное упорядоченное состояние. Согласно представлениям данной работы синтез атомоподобных квазичастиц на основе структурных элементов предыдущего уровня осуществляется через механизм образования псевдомолекулярных комплексов атомных частиц. Формирование этих комплексов в процессе понижения температуры жидкости сопровождается релаксацией их локальной структуры к псевдоатомному состоянию. В свою очередь, псевдоатомное состояние квазичастиц является исходным для формирования на их основе псевдомолекулярных квазичастиц следующего уровня сложности. Реализация такого рода "*атомно-молекулярных*" состояний определяет формирование внутренних степеней свободы композитных образований, отвечающих движению структурных элементов атомно-молекулярных квазичастиц относительно друг друга. Такие степени свободы мы будем называть "псевдоатомными" степенями свободы.

Логика образования структурных элементов квантовой жидкости в данной работе основана на сходстве принципов и механизмов организации композитных квазичастиц гелия-4 с принципами и механизмами образования внешних электронных оболочек молекул и атомы. В дальнейшем для краткости этот принцип подобия механизмов организации составных квазичастиц жидкости механизмами образования молекул и атомов мы будем называть "*принцип АМ-подобия структуры жидкости*"[2].

---

[2] В этой связи, при описании механизмов установления связей между структурными элементами жидкости в процессе формирования композитных квазичастиц, мы будем формально пользоваться терминологией и обозначениями, которые используются для описания и обозначения электронных состояний атомов и межатомных связей в физике атомов и молекул.

В первой части работы изложены основные идеи микроскопической физики жидкого гелия[3]. Во второй части работы эти идеи использованы для расчета теплоемкости гелия-4 в окрестности $\lambda$-перехода системы в когерентное состояние.

## Part 1. Микроскопическая физика генерации фермионных свойств макроскопического ансамбля атомов гелия-4 и их эволюции при понижении температуры.

> *"Очень часто упрощенная модель проливает больше света на то, как устроена природа явления, чем любое число вычислений "ab initio"…"*
>
> **Ф. Андерсон,** Нобелевские лекции по физике, 1977г.

### 1.1 Взаимодействующие атомы гелия-4 как фермионные псевдоатомы. Гиротоны.

#### 1.1.1 *Вихревое поле и наведенные спины атомов. Формирование атомоподобных квазичастиц.*

Эффект межатомного взаимодействия макроскопического ансамбля $N_0$ атомов гелия в объеме $V$ связывается в данной статье с формированием внутреннего вихревого поля, создаваемого электронными и ядерными зарядами макросистемы, движущимися относительно друг друга. С этой точки зрения атомы гелия погружены во внутреннее вихревое поле, которое определяет эффект трансмутации ансамбля взаимодействующих бозонных атомов в систему свободных фермионных квазичастиц.

Окружающую каждый атом область пространства, объем которого равен удельному объему макросистемы $\upsilon = V/N_0$, можно рассматривать как "элементарную ячейку" квазистационарной локализации атома. Отклик атома на вихревое поле зависит от температуры системы и в общем случае может быть охарактеризована двумя взаимообусловленными эффектами: во-первых, колебаниями электронных орбиталей атомных электронов и, во-вторых, квазизамкнутым движением (осцилляциями) атома гелия в окрестности центра "элементарной ячейки". При этом второй эффект в явном виде может проявляться только в конденсированном состоянии вещества.

С точки зрения полуклассического представления колебания электронных орбиталей представляют собой нестационарную прецессию плоскостей орбитального движения атомных электронов. Такой вид "дополнительного" движения электронной подсистемы атома в поле вихря можно рассматривать как эффект, порождающий локализованный на атоме "*мнимый электрон*". "Рождение" мнимого электрона есть результат возбуждения электронной подсистемы атома. Как следствие, статистика реализованных квантовых состояний частиц, обусловленных такого рода возбуждениями, должна отвечать статистике фермионов со спином, определяемым квантовым числом $s=1/2$ (в дальнейшем величину углового момента будем определять значением соответствующего квантового числа). В таком случае, мнимый электрон, локализованный на нейтральном атоме гелия, обуславливает "приобретение" атомом гелия нового качества, которое мы определим, как "*наведенный спин*" (или "*псевдоспин*") с угловым моментом $s=1/2$. И, тем самым, "превращает" бозон в фермионную квазичастицу. С этой точки зрения, мнимый электрон является носителем псевдоспинового состояния атома, обусловленного взаимодействием атома с вихревым полем. Атом гелия с локализованным на нем наведенным спином представляет собой фермионную квазичастицу, которую будем называть *"гиротон"*. Таким образом, определяя вихревое поле, как внутреннее свойство атомной частицы, мы переходим от анализа поведения ансамбля взаимодействующих атомов гелия-4 к рассмотрению ансамбля "свободных" фермионных квазичастиц.

---

[3] Для сокращения объема статьи и ограничения количества вновь вводимых представлений в настоящей работе будут упрощены или опущены те положения и детали микроскопической структуры квазичастиц, которые не используются при расчете теплоемкости в окрестности $\lambda$-перехода в явном виде.

Согласно вышесказанному, гиротон в области температур $T<T_{cr}$, где $T_{cr}$ отвечает критической температуре перехода гелия-4 из газообразного в жидкое состояние, совершает квазизамкнутое движение в области центра своей квазистационарной локализации (помимо других движений со значительно большими характерными временами, которые здесь мы не рассматриваем). Такое динамическое образование, заключенное в соответствующей "элементарной ячейке", можно рассматривать как гиротонный псевдоатом. Эту атомоподобную фермионную квазичастицу мы будем называть "*гироатом*". В таком представлении гиротоны играют роль псевдоатомных электронов с эффективной массой равной массе атома гелия $M_{He}$. Поэтому гиротоны в соответствующих случаях будем называть псевдоэлектронами. При этом собственно гироатомы обладают как псевдоспиновыми, так и псевдоатомными степенями свободы[4].

Поведение фермионных гиротонов в среде подобных же частиц зависит от температуры и определяется структурой и квантовым состоянием образуемых на их основе композитных гироатомов (атомно-молекулярных квазичастиц). Образование атомно-молекулярных квазичастиц определяет принципиальную возможность разделения их движения по псевдоатомным и псевдоспиновым степеням свободы.

### 1.1.2 *Условия и форма реализации фермионных свойств бозонных атомов.*

Механизм бозе-ферми трансмутации статистических свойств атомов гелия-4 связан с ограничением свободы их поступательного движения в поле $\Phi$ потенциального межатомного взаимодействия. В результате, часть тепловой энергии поступательного движения атомов "преобразуется" во внутриатомную энергию псевдоспиновых степеней свободы гиротонных квазичастиц. При этом энергия всех реализованных псевдоспиновых степеней свободы оказывается полностью "компенсирована" отрицательной энергией взаимодействия атомов, "затрачиваемой" на изменение характера движения атомов. То есть, для "приобретения" $N_a$ атомами гелия спиновых степеней свободы и, тем самым, для образования $N_e=N_a$ гиротонных квазичастиц необходимо, чтобы отрицательная потенциальная энергия взаимодействия всех $N_0$ атомов ансамбля сравнялась по абсолютной величине с энергией теплового поступательного движения $N_a$ степеней свободы атомов при температуре $T$:

$$|\varphi(\upsilon)| = (N_e/N_0) \cdot (1/2) k_B T \qquad (1)$$

здесь $\upsilon = \upsilon(T)$ – зависящий от температуры удельный объем атомной системы; $|\varphi(\upsilon)|$ - модуль потенциальной энергии взаимодействия атомов гелия в расчете на одну частицу.

---

[4] В дальнейшем изложении мы иногда будем опускать в словах приставку "псевдо" или "квази" в случаях, когда будет очевидно, что речь идет о структурно-динамических элементах гироатомных квазичастиц.

С позиции представления гелия как ансамбля псевдоатомных квазичастиц и в соответствии с оговоренным ранее принципом АМ-подобия структуры жидкости (см. Введение, стр. 4), наведенные спины "порождают" гироатомные псевдоэлектроны, которые занимают состояние $1s$ псевдоэлектронной оболочки соответствующего гироатома. Взаимодействие между атомами ведет к расщеплению $1s$ уровня псевдоэлектронной оболочки и формированию зоны энергетических состояний гиротонов. Энергетическая ширина такой зоны равна энергии $|\varphi(\upsilon)|$ и растет вместе с числом индуцированных мнимых электронов по мере понижения температуры системы. При этом, согласно (1), внутренняя энергия псевдоспиновых состояний атомов полностью компенсирует энергию межатомного взаимодействия до тех пор, пока при некоторой характеристической температуре $T_0$ не будут индуцированы все возможные мнимые электроны. Поэтому можно полагать, что энергетические состояния гироатомных псевдоэлектронов отвечают энергетическим уровням абсолютно вырожденной системы свободных фермионов с эффективной температурой $T_{eff}=0K$, массой $M=M_{He}$ и спином $s=1/2$. И это условие сохраняется до тех пор, пока модуль отрицательной энергии межатомного взаимодействия не достигнет своего максимального значения $|\varphi_0|$.

Полное число мнимых электронов $N_{e0}$, которое может быть реализовано в системе из $N_0$ атомов гелия, равно числу атомных электронов $2N_0$. Отсюда следует, что при понижении температуры системы процесс генерации наведенных спинов и, следовательно, образования гироатомных псевдоэлектронов, включает в себя две стадии. Первая стадия соответствует процессу рождения $N_e=N_0=(1/2)N_{e0}$ мнимых электронов, когда все $N_0$ частиц атомного ансамбля "приобретают" в поле вихря псевдоспиновые степени свободы и, тем самым преобразуются в фермионные квазичастицы. При этом движение каждого атома гелия по одной из трех поступательных степеней свободы оказывается "вымороженным". С позиции представлений данной работы это соответствует условиям достижения системой критических термодинамических параметров $T_{cr}$, $V_{cr}$ и $P_{cr}$ перехода системы из газового в жидкое состояние. Соответствующие псевдоэлектроны будем называть первичными псевдоэлектронами или $K$-псевдоэлектронами. $K$-псевдоэлектроны реализуются при $T \geq T_{cr}$ и постепенно заполняют состояния $K$-зоны по мере понижения температуры. Эти состояния отвечают основному состоянию свободных фермионов с $T_{eff}=0K$. Полное число $K$-псевдоэлектронов при температуре $T=T_{cr}$ равно $N_{K0}=(1/2)N_{e0}=N_0$ и все атомы системы приобретают свойства фермионов. Согласно условию (1) граничная энергия $K$-псевдоэлектронов при $T=T_{cr}$ отвечает энергии Ферми $E_F(N_{K0})=k_B T_{cr}/2$:

$$E_{F_{cr}}(N_{K0}) = \frac{1}{2}k_B T_{cr} = \frac{\hbar^2}{2M_{He}} \cdot \left(\frac{6\pi^2 N_{K0}}{g_s \cdot V_{cr}}\right)^{2/3} = \frac{\hbar^2}{2M_{He}} \cdot \left(\frac{3\pi^2}{\upsilon_{cr}}\right)^{2/3} \qquad (2)$$

Здесь фактор вырождения $g_s=2s+1$ соответствует спиновому состоянию псевдоэлектронов $s=1/2$; $V_{cr}=V(T_{cr})$ и $\upsilon_{cr}=V_{cr}/N_0$ – критические объем и удельный объем гелия-4 при $T=T_{cr}$.

## 1.2 Первый этап конденсации квазичастиц в гелии. Конденсация первичных $K$-псевдоэлектронов и формирование ферми-жидкости двухчастичных гироатомов.

### 1.2.1 *Механизм образования парно-коррелированных состояний гиротонов и формирование подсистемы "вторичных" псевдоэлектронов.*

С позиции квазичастичного представления заполненными при температуре $T=T_{cr}$ являются только половина из максимально возможного числа $N_{e0}$ состояний зоны псевдоэлектронных квазичастиц (см. подраздел 1.1.2). Модуль потенциальной энергии взаимодействия атомов в расчете на одну частицу при $T<T_{cr}$ становится больше энергии внутреннего движения независимых спиновых степеней свободы первичных псевдоэлектронов. При этом возможности одного независимого атома скомпенсировать "избыточную" потенциальную энергию путем реализации свободного псевдоспинового состояния оказываются исчерпанными. Поэтому система становится неустойчивой по отношению к формированию новых структурных единиц макросистемы путем установления парно-корреляционных связей между гироатомами.

Механизм установления парных корреляций гироатомов можно связать с процессом *s*-спаривания спинов тех *K*-псевдоэлектронов, которые находятся на поверхности Ферми при текущей температуре $T \leq T_{cr}$. Это означает "конденсацию" *Ks*-псевдоэлектронных состояний на поверхности Ферми неспаренных *K*-псевдоэлектронов. При этом спаренные *K*-псевдоэлектроны теряют свойство индивидуальных фермионных квазичастиц. Данный процесс есть проявление эффекта фазового переход гелия из газообразного в жидкое состояние.

Гироатомные связи, обусловленные спариванием *K*-псевдоэлектронов, будем называть первичными связями, а парно-коррелированные состояния фермионных гиротонов будем называть "*первичными парами*". Первичная пара гироатомных квазичастиц представляет собой виртуальное образование, которое является проявлением существования некоторого параметра ближнего порядка, характеризующего моноатомную макросистему в жидком агрегатном состоянии.

При понижении температуры $T<T_{cr}$ и установлении первичных парных связей уменьшается удельный объем $\upsilon(T)$ жидкости и растет степень ограничения свободы поступательного движения индивидуальных атомов. Как следствие, реализуются условия для второй стадии процесса генерации мнимых электронов и, тем самым, для образования "*вторичных*" псевдоэлектронов. Согласно предыдущему разделу данный процесс идет в интервале температур $T_0 \leq T < T_{cr}$.

Согласно вышесказанному, конденсация *K*-псевдоэлектронов сопровождается генерацией вторичных наведенных спинов атомов гелия вплоть до температуры $T_0$. При этом общее число первичных и вторичных свободных псевдоэлектронов в интервале температур $T_0 \leq T < T_{cr}$ остается равным $N_0$. Следовательно, каждый гироатом, участвующий в парно-корреляционном взаимодействии, несет по одному неспаренному вторичному псевдоэлектрону, реализованному при $T<T_{cr}$. То есть атомы гелия, как индивидуальные квазичастицы сохраняют статус гиротонных фермионов в области температур $T_0 \leq T < T_{cr}$. Для частиц, образующих первичные пары этот статус обеспечивается вторичными наведенными спинами. При этом энергетические состояния вторичных псевдоэлектронов, как и в случае реализации *K*-псевдоэлектронов, отвечают состояниям фермионов с эффективной температурой $T_{eff}=0K$, массой $M=M_{He}$ и спином $s=1/2$.

Таким образом, при переходе системы в состояние жидкости, с одной стороны, формируются первичные пары гироатомов. С другой стороны, в той же области температур $T_0 \leq T < T_{cr}$ формируются вторичные гиротонные псевдоэлектроны, чьи энергетические состояния отвечают состояниям свободных фермионов. Причем оба процесса происходят синхронно. Отсюда можно заключить, что вторичные псевдоэлектроны реализуются попарно и, соответственно, попарно заполняют состояния абсолютно вырожденной системы фермионов внутри сферы Ферми. Это означает, что вторичные псевдоэлектроны первичных пар имеют противоположно направленные спины. Наличие парных корреляций между частицами, несущими свободные спины абсолютно вырожденной системы фермионов можно рассматривать, как эффект формирования основного состояния гиротонной ферми-жидкости.

Двойственная природа парно-коррелированных гироатомов при $T < T_{cr}$ обусловливает формирование у них общих псевдоатомных и псевдоспиновых степеней свободы. Это позволяет рассматривать первичные пары гироатомов как двухчастичные псевдоатомы, которые мы будем называть "*бигиротоны*". Заметим, что на стадии формирования фермионных свойств атомной системы явного разделения движения гироатомов по псевдоатомным и псевдоспиновым степеням свободы при $T \geq T_0$ не происходит, поскольку для этого необходимо, чтобы энергия взаимодействия атомов была больше кинетической энергии псевдоэлектронных состояний гиротонов. Поэтому, вплоть до температуры $T_0$ завершения формирования подсистемы вторичных псевдоэлектронов, их энергетические состояния отвечают состояниям свободных фермионов.

С позиции принципа АМ-подобия структуры жидкости (см. Введение, стр. 4), вторичные псевдоэлектроны первичных пар занимают состояния в $L$- псевдооболочке бигиротонных псевдоатомов. Поэтому вторичные псевдоэлектроны будем называть "$L$-псевдоэлектроны". И эти $L$-псевдоэлектроны попарно заполняют состояния с противоположными спинами в "своей" (вторичной) $L$-зоне, энергетические уровни которой соответствуют энергетическим уровням абсолютно вырожденной системы фермионов. Полное число псевдоэлектронных состояний, реализованных при текущей температуре $T_0 \leq T < T_{cr}$, равно $N_e = N_{K0}+N_L$. При этом число $N_L$ "рожденных" $L$-псевдоэлектронных квазичастиц равно, очевидно, числу $N_{K-}$ конденсированных $K$-псевдоэлектронов. Отсчет энергии $L$-псевдоэлектронов, заполняющих $L$-зону, ведется от принимаемой за ноль энергии верхнего заполненного при текущей температуре уровня $K$-зоны. При температуре $T < T_{cr}$ эта зона заполнена $N_K = (N_{K0} - N_{K-}) = (N_0 - N_L)$ непарными $K$-псевдоэлектронами. Таким образом, $L$-псевдоэлектроны при температуре $T$ заполняют состояния свободных фермионов с $T_{eff}=0K$ и химическим потенциалом $\mu(N_L)_{T \geq T_0}$, равным энергии Ферми $E_{F_L}(N_L)$:

$$E_{F_L}(N_L) = \frac{N_e}{N_0} \cdot \frac{1}{2} k_B T - \frac{(N_0 - N_L)}{N_0} \cdot \frac{1}{2} k_B T = \frac{N_L}{N_0} k_B T = \frac{\hbar^2}{2M_{He}} \cdot \left( \frac{6\pi^2}{g_s} \cdot \frac{N_L/N_0}{\upsilon(N_e)} \right)^{2/3} \quad (3)$$

Здесь удельный объем псевдоэлектронной системы $\upsilon(N_e)$ определяется полным числом реализованных псевдоэлектронных состояний $N_e$ при постоянном исходном (в плане реализации полного числа псевдоэлектронов) объеме жидкости $V_{cr} = V(T_{cr})$:

$$\upsilon(N_e) = \frac{V_{cr}}{N_e(T)} = \frac{\upsilon_{cr}}{1 + N_L/N_0} \quad (4)$$

Химический потенциал $\mu_0$ и соответствующая ему энергия Ферми $E_{F0}(N_{L0})$ системы $L$-псевдоэлектронов при температуре $T_0$ реализации всех $N_{L0}=N_0$ вторичных псевдоэлектронов, согласно (1), (3) и (4), равны модулю энергии межатомного взаимодействия в расчете на одну частицу:

$$\mu_0 = E_{F0}(N_{L0}) = -\varphi_0 = k_B T_0 = \frac{\hbar^2 k_0^2}{2M_{He}} = \frac{\hbar^2}{2M_{He}} \cdot \left( \frac{3\pi^2 N_0}{V_0} \right)^{2/3} = \frac{\hbar^2}{2M_{He}} \cdot \left( \frac{3\pi^2}{\upsilon_0} \right)^{2/3} \quad (5)$$

Здесь $\varphi_0(\upsilon_0) = -k_B T_0$ соответствует максимальной отрицательной потенциальной энергии парного взаимодействия атомов в расчете на одну частицу (с точностью до учета внутренней энергии $Ks$-спаривания первичных пар); $V_0 = V(T_0)$ – объем атомной системы при температуре $T_0$; $\upsilon(T_0) = V_0/N_0 = \upsilon_0$ отвечает удельному объему гелия при $T=T_0$; $k_0$ есть граничный волновой вектор полностью заполненной $L$-псевдоэлектронами сферы Ферми при $T=T_0$.

### 1.2.2 *Разделение состояний псевдоэлектронной жидкости на состояния квазичастиц Ls- и Lp-типов.*

Согласно подразделу 1.2.1 вторичные псевдоэлектроны при температуре $T_0$ заполняют состояния $L$-оболочки всех бигиротонных псевдоатомов. В свою очередь, энергетические состояния $L$-оболочки подразделяются на состояния, отвечающие орбитальным моментам $l=0$ ($s$-подоболочка) и $l=1$ ($p$-подоболочка). Поэтому расщепление состояний $L$-оболочки, которое происходит в результате межатомного взаимодействия, приводит к образованию $L$-зоны, состоящей из $Ls$ и $Lp$ подзон. Согласно предыдущему подразделу $L$-зона при температуре $T_0$ заполнена $N_0$ вторичными псевдоэлектронами с эффективной температурой $T_{eff}=0K$ (ф-ла (5)). В таком случае, псевдоэлектроны внутри сферы Ферми радиуса $k_0$ разделяются на две подсистемы (см. рис. 1). Первая подсистема включает $L$-псевдоэлектроны с энергией относительного движения атомных частиц первичной пары ниже энергии $\varepsilon_1$, определяемой орбитальным квантовым числом $l=1$. И, соответственно, вторую подсистему образуют $L$-псевдоэлектроны с энергией $\varepsilon \geq \varepsilon_1$. Соответствующие подсистемы частиц будем называть $Ls$- и $Lp$- компонентами псевдоэлектронов. Квазичастицы двух подсистем, заполняющие соответствующие подзоны, обозначим как частицы $Ls$- и $Lp$-типов.

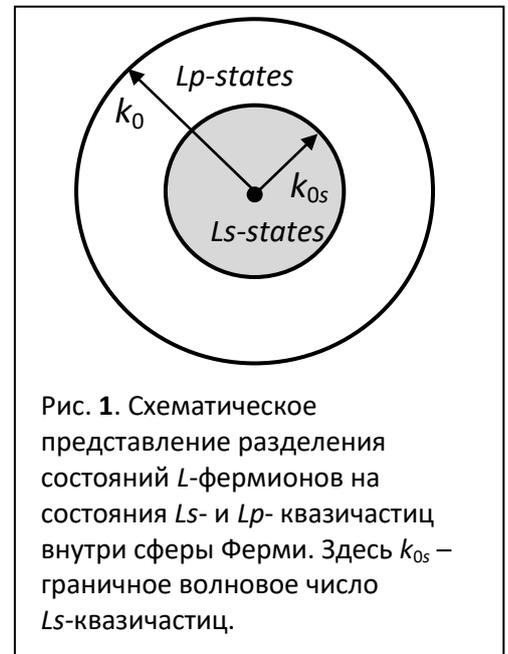

Рис. **1**. Схематическое представление разделения состояний $L$-фермионов на состояния $Ls$- и $Lp$- квазичастиц внутри сферы Ферми. Здесь $k_{0s}$ – граничное волновое число $Ls$-квазичастиц.

С одной стороны, при $T=T_0$ все вторичные псевдоэлектроны представляют собой свободные фермионы со спинами $s=1/2$. С другой стороны, $Lp$-псевдоэлектроны можно рассматривать как самостоятельную подсистему парно-коррелированных квазичастиц, которые обладают (дополнительным к спиновому моменту $s=1/2$) орбитальным моментом $l=1$. При этом $Lp$-квазичастицы таких пар с равной вероятностью могут находиться в состояниях, отвечающих квантовым числам $j=1/2$ или $j=3/2$. В данном представлении степень вырождения $Lp$-частиц оказывается в два раза больше, чем степень вырождения $Ls$-частиц. В таком случае, относительные доли частиц $Ls$- и $Lp$-типов могут быть определены из условия равенства удельных термодинамических потенциалов двух гипотетических фермионных подсистем со степенью вырождения квантовых состояний $g_{1/2}=2$ и $g_{3/2}=4$:

$$\frac{N_{0s}}{N_0}\cdot\frac{\hbar^2}{2M_{He}}\cdot\left(\frac{6\pi^2}{g_{1/2}\cdot\upsilon_0}\right)^{2/3} = \frac{N_{0p}}{N_0}\cdot\frac{\hbar^2}{2M_{He}}\cdot\left(\frac{6\pi^2}{g_{3/2}\cdot\upsilon_0}\right)^{2/3} \quad (6)$$

Значения $N_{0s}$ и $N_{0p}$ отвечают числам $Ls$- и $Lp$-частиц соответственно: $N_{0s}+N_{0p}=N_0$. Относительные доли частиц $N_{0s}/N_0$ и $N_{0p}/N_0$ определяют вероятность нахождения фермионных квазичастиц в $Ls$- и $Lp$-состояниях.

Из (6) получаем, что относительные доли $Ls$- и $Lp$-частиц равны[5]:

---

[5] С позиции квазичастичного представления тот же результат может быть получен с использованием канонического распределения по псевдоатомным степеням свободы гиротонов. Такой подход предполагается изложить в другой работе.

$$N_{0s}/N_0 = \left(1+2^{2/3}\right)^{-1} = 0.3865;$$
$$N_{0p}/N_0 = \left(1 - N_{0s}/N_0\right) = 0.6135 \tag{7}$$

Эффекты парно-коррелированного состояния фермионных квазичастиц и их разделение по *s*- и *p*-состояниям есть проявление свойств ферми-жидкости.

### 1.3 Второй этап конденсации квазичастиц в гелии. бозе-конденсация гиротонных квазичастиц и образование фракции композитных бозонов в гелии-I.

#### 1.3.1 *Механизм формирования и структура композитных квазичастиц конденсатной фракции гелия-I.*

С позиции "канонического" представления температура перехода газа свободных бозонов в бозе-конденсатное состояние $T=T_{BK}$ определяется условием $\mu$=0, где $\mu$ - химический потенциал системы. При температуре $T<T_{BK}$ система бозе-частиц разделена на подсистемы частиц надконденсатной и конденсатной фракций, которые находятся в равновесии с равными нулю химическими потенциалами. Переход в бозе-конденсатное состояние системы атомных частиц при помещении их во внешнее поле $\varPhi$ определяется условием $\mu^{(tot)} = 0$, где $\mu^{(tot)}$ есть полный удельный термодинамический потенциал системы:

$$\mu^{(tot)} = \mu + \varphi = 0 \tag{8}$$

Здесь $\varphi$ - потенциальная энергия системы в поле $\varPhi$ в расчете на одну частицу. Условие (8), очевидно, должно оставаться в силе и для системы свободных фермионов, претерпевающей переход в бозе-конденсатное состояние.

Согласно введенным нами представлениям, *L*-псевдофермионы (гиротоны) в области температур $T_0 \leq T < T_{cr}$ ставятся в соответствие атомам гелия, которые находятся в поле $\varPhi$ межатомного взаимодействия. Потенциальная энергия атомов, "обеспечивающая" образование $N_e$ гиротонных квазичастиц, определяется ур. (1). Поле $\varPhi$ является внешним полем по отношению к системе свободных фермионных гиротонов. С этой точки зрения, полный удельный термодинамический потенциал $\mu^{(tot)}$ гиротонной системы *L*-псевдоэлектронов с химическим потенциалом $\mu(N_L)_{T \geq T_0}$, равным энергии Ферми $E_{F_L}(N_L)$, определяется соотношением (см. ур. 1, 3)):

$$\mu^{(tot)}(N_L)_{T \geq T_0} = \frac{N_L}{N_0} \cdot k_B T - \frac{N_e}{N_0} \cdot (1/2) k_B T = -\frac{N_0 - N_L}{N_0} \cdot (1/2) k_B T \tag{9}$$

При $T=T_0$, согласно подразделу 1.2.1, $N_L=N_{L0}=N_0$. Таким образом, значение полного удельного термодинамического потенциала $\mu^{(tot)}_{T=T_0}(N_{L0})$ (ур. (8)) при $T=T_0$ равно нулю. Следовательно, температура $T_0$ отвечает температуре перехода системы гиротонов в бозе-конденсатное состояние.

Все *L*-псевдоэлектроны бигиротонных псевдоатомов при температуре $T_0$ занимают попарно состояния абсолютно вырожденной системы фермионов. Энергия $k_B T$ спинового движения тех *L*-псевдофермионов, которые находятся на поверхности Ферми при текущей температуре $T<T_0$, становится меньше энергии межатомного взаимодействия $|\varphi_0(\upsilon_0)| = k_B T_0$ по абсолютной величине. Это означает, что спиновые состояния соответствующих бигиротонов перестают быть независимыми. Как следствие, между бигиротонами начинают формироваться новые (вторичные) парные связи путем спаривания $L$-псевдоэлектронов, входящих в состав двух соседних первичных пар. Причем, в образовании парных состояний $L$-псевдоэлектронов вплоть до некоторой характеристической температуры $T_c$ принимают участие только частицы $Lp$-типа (см. рис. 1). В результате два гиротона, находящиеся в разных (соседних) первичных парах, образуют вторичные $p$-пары. Тем самым происходит спаривание двух первичных пар и образование парно-парных состояний гиротонов. При этом $L$-псевдоэлектроны в первичных парах также теряют взаимную независимость. С этой точки зрения можно говорить о том, что при $T<T_0$ между гиротонами каждой первичной пары также устанавливаются вторичные связи путем $s$-спаривания их $L$-псевдоэлектронов. Это означает, что парно-парные состояния гиротонов определяются суперпозицией состояний $s$- и $p$- псевдоэлектронных пар. С позиции принципа АМ-подобия структуры жидкости (см. стр. 4) эти процессы приводят к формированию атомно-молекулярной квазичастицы с общей для четырех гиротонов атомоподобной $Lp$-подоболочкой. Такие квазичастицы мы будем обозначать, как "*бигиротонные дублеты*" или просто "*дублеты*".

Дублеты представляют собой композитные бозоны. Как следствие, при $T< T_0$ начинается процесс разделения системы на надконденсатную фракцию $N_0-N_c=N_{nc}$ фермионов и конденсатную фракцию бозонов, формируемых на базе $N_c$ гиротонов. Данный этап конденсации реализуется в температурном интервале $T_c \leq T < T_0$ до тех пор, пока все $L$-псевдоэлектроны $p$-типа не образуют спаренные вторичные пары. И, тем самым, перейдут в бозе-конденсатное состояние в форме бигиротонных дублетов. Образование бозе-конденсатной фракции бигиротонных дублетов обусловливает эффект неявного разделения движения гироатомных квазичастиц по псевдоспиновым и псевдоатомным степеням свободы.

В структурном отношении конденсатные дублеты представляют собой симплексы Делоне [5, 6] в виде тетраэдров. Симплексы Делоне являются основными структурными элементами некогерентной конденсатной фракции гелия-I.

### 1.3.2 *Температура разделения системы парно-коррелированных гиротонов на бозе-конденсатную и фермионную надконденсатную фракции.*

С точки зрения термодинамики переход системы в бозе-конденсатное состояние неотличим от процесса "классического" перехода газ-жидкость [7]. Следовательно, процесс бозе-конденсации фермионных квазичастиц гелия-4 должен протекать вдоль линии сосуществования газового и жидкого агрегатных состояний гелия $P=P_{SVP}(T)$, где $P_{SVP}$ – давление насыщенных паров.

На рис. 2 приведена зависимость удельного объема жидкого гелия-4 от температуры при давлении насыщенных паров, которая построена на основании рекомендованных в работе [8] данных по плотности жидкого гелия. На том же рисунке нанесена зависимость удельного объема $\upsilon$ абсолютно вырожденной системы частиц с массой $M=M_{He}$ и спином $s=1/2$ от температуры Ферми $T_F$. Согласно соотношениям (1), (3-5), температура $T_0$ гелия в точке фазового перехода

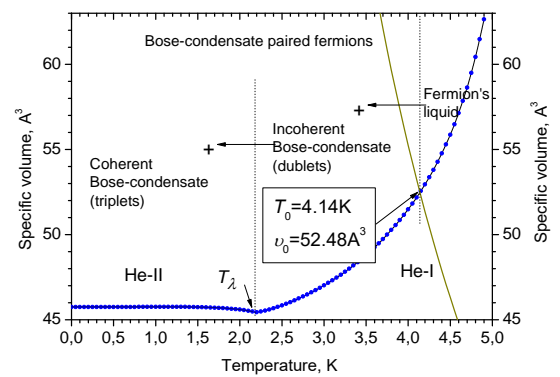

Рис. 2. Определение температуры бозе-конденсации гиротонных фермионов. Экспериментальная зависимость удельного объема жидкого гелия от температуры показана сплошными кружочками, соединенными линией. Связь между удельным объемом и температурой Ферми системы фермионов (ур.( 5)) показана сплошной линией. Точка пересечения определяет термодинамические параметры жидкого гелия, при которых реализуются условия бозонизации гиротонной системы.

фермионных гиротонов в состояние бозе-конденсата равна температуре Ферми $T_F$ системы $N_0$ вторичных псевдоэлектронов. Из этого условия получаем значения температуры $T_0$ и соответствующего ей удельного объема $\upsilon_0$:

$$T_0 = 4.14 K$$
$$\upsilon_0 = 52.48 \text{Å}^3 \qquad (10)$$

Результат (10) указывает на то, что процесс бозе-конденсации в гелии-I и переход системы в квантово-когерентное состояние относятся разным фазовым переходам.

Согласно формулам (2), (4) и (5) температура $T_0$ перехода ансамбля гиротонных квазичастиц в бозе-конденсатное состояние связана с критической температурой гелия-4 соотношением $T_0 = 2^{-1/3} T_{cr}$. Подставляя значение $T_0 = 4.14 K$ в это соотношение получаем величину, практически точно совпадающую с экспериментальным значением критической температуры гелия-4 (см. например, [8]):

$$T_{cr} = 2^{1/3} T_0 \approx 5.22 K \qquad (11)$$

Полученный результат подтверждает правомерность изложенных выше модельных представлений о микроскопических процессах, протекающих в гелии-4 при понижении температуры.

### 1.4 Третий этап конденсации гиротонных квазичастиц в гелии. Переход гелия-I в когерентное состояние гелий-II. Атомно-молекулярная жидкость трехпарных гиротонных квазичастиц.

#### 1.4.1 *Механизм разделения конденсатной фракции гелия-II на когерентную и некогерентную подсистемы. Структура композитных квазичастиц когерентной составляющей конденсатной фракции.*

При температуре $T<T_c$ бигиротоны надконденсатной фракции несут $L$-псевдоэлектроны только $Ls$-типа. В соответствии с логикой развиваемых представлений процесс бозонизации системы при $T<T_c$ должен продолжаться путем спаривания тех бигиротонных псевдоатомов, $L$-псевдоэлектроны которых находятся на поверхности Ферми при текущей температуре $T$. С точки зрения псевдоатомного бигиротонного представления, соответствующие первичные пары находятся в синглетном состоянии с завершенными псевдоатомными $Ls$ подоболочками. В этом случае, установление парных связей между двумя бигиротонами путем образования гибридных $sp$ псевдомолекулярных орбиталей двух возбужденных бигиротонов $Ls$-типа оказывается энергетически невыгодным. Однако, бозонизация фермионных состояний гиротонов возможна при посредничестве ранее сформированных дублетов бозе-конденсатной фракции гелия-I.

Доля гиротонов в дублетной бозе-конденсатной фракции при температуре $T=T_c$ составляет, грубо округляя, 2/3 всех частиц (см. результаты расчетов (7)). И, соответственно, число фермионных гиротонов надконденсатной фракции составляет приблизительно 1/3 всех гиротонных квазичастиц. Это позволяет рассматривать конденсатную фракцию бигиротонных дублетов как некоторую "матрицу", в которой бигиротоны надконденсатной фракции "растворены" в соотношении примерно 1:1. Таким образом, в равновесном состоянии каждый надконденсатный бигиротонный псевдоатом при температуре $T=T_c$ окружен атомоподобными дублетами конденсатной фракции. И наоборот, ближайшими соседями дублетных псевдоатомов являются надконденсатные бигиротоны. В таком случае, при температуре $T \approx T_c$ области корреляционного взаимодействия конденсатных квазичастиц перекрываются во всем объеме. Это приводит к бозонизации надконденсатных квазичастиц $Ls$-типа при понижении температуры путем формирования когерентно связанных между собой новых квазичастичных структурных элементов жидкого гелия.

С позиции принципа АМ-подобия структуры жидкости (см. стр. 4) механизм бозонизации надконденсатных квазичастиц $Ls$-типа в гелии-II связан с эффектом $sp$ гибридизации и объединения псевдоэлектронных орбиталей надконденсатного бигиротона и орбиталей соседнего с ним конденсатного дублета. Как результат, на основе бигиротонных синглета и дублета формируются атомно-молекулярные квазичастицы в форме парно-коррелированных гиротонных троек. Такие квазичастичные структуры будем называть "*бигиротонные триплеты*" или просто "*триплеты*". При этом незамкнутость псевдоэлектронной $L$-оболочки атомоподобного состояния триплетных квазичастиц определяет установление корреляционной связи между триплетами и, тем самым, формирование квазиполимерных цепочек на их основе[6].

Конденсатная фракция $N_c$ частиц гелия-II разделяется на две подсистемы за счет формирования новых структурных единиц жидкости. Первая подсистема состоит из "синтезированных" ранее в фазе I бигиротонных дублетов, образованных $N_{b2}=N_c–3(N_c-N_{0p})$ гиротонными квазичастицами. Вторая подсистема состоит из "синтезированных" в фазе II когерентно-связанных бигиротонных триплетов, которые образованы $N_{b3}=3(N_c-N_{0p})$ гиротонами путем объединения $(N_c–N_{0p})$ $Ls$- частиц с $2(N_c–N_{0p})$ конденсатными $Lp$-частицами. Такое разделение конденсатной фракции на две составляющие обуславливает скачок теплоемкости системы при температуре $T=T_c$ (см. Part 2, подраздел 2.3.3, пункт $d$). Это означает, что при $T=T_c$ в системе происходит фазовый переход второго рода в упорядоченное состояние синглетов и дублетов относительно друг друга в рамках триплетных структур. Степень порядка $\chi$ в системе в целом возрастает от 0 до 1 по мере роста числа триплетных квазичастиц. Таким образом, температура $T_c$ есть не что иное, как температура $T_\lambda$ перехода жидкого гелия-4 из состояния "гелий-I" в когерентное упорядоченное состояние "гелий-II".

### 1.4.2 *Температура перехода гелия-I в когерентное состояние гелий-II.*

Число реализованных квазинезависимых $L$-псевдоэлектронов при температуре $T=T_0$ становится равным числу частиц атомного ансамбля. При дальнейшем понижении температуры происходит уменьшение числа свободных $L$-псевдоэлектронов в результате их вторичного спаривания и конденсации на уровне Ферми $L$-зоны, соответствующем энергии парной связи $\varphi_0$. Поэтому внутри сферы Ферми радиуса $k_0$ (ур. (5)) при температуре $T<T_0$ появляются незаполненные уровни энергетических состояний. Это означает, что эффективная температура фермионов становится отличной от нуля.

Переход $N_c=N_0– N_{nc}$ фермионов в бозе-конденсатное состояние композитных бозонов исключает соответствующие атомы гелия из процесса формирования квазинезависимых псевдоспиновых состояний $L$-псевдоэлектронов (в силу насыщенного характера связей между структурными элементами атомно-молекулярных квазичастиц). Поэтому реализация $N_{nc} = N_L$ надконденсатных $L$-псевдоэлектронных состояний в области температур $T<T_0$ определяется модулем энергии взаимодействия $|\varphi_{nc}|$ только между $N_{nc}$ надконденсатными атомами гелия:

$$|\varphi_{nc}| = (N_L/N_{nc}) \cdot k_B T = k_B T \tag{12}$$

---

[6] В рамках развиваемой логики возможны несколько вариантов бозонизации $Ls$-квазичастиц при посредничестве бигиротонных дублетов конденсатной фракции. Однако все они приводят, по сути, к одному результату образования триплетов бигиротонных квазичастиц. Поэтому мы не будем подробно рассматривать особенности формирования структуры композитных квазичастиц упорядоченной составляющей конденсатной фракции гелия-II в данной работе.

Это означает, что внешнее поле индивидуально взаимодействующих атомов гелия, в котором находятся надконденсатные псевдофермионы, определяется взаимодействием только между собственно $N_{nc}$ атомами. При этом условием равновесия надконденсатной и конденсатной подсистем является равенство нулю полного удельного термодинамического потенциала надконденсатной фракции фермионов при $T<T_0$:

$$\mu_{nc} + \varphi_{nc} = 0 \tag{13}$$

Соответственно, химический потенциал фермионов в области температур $T<T_0$ равен модулю энергии взаимодействия надконденсатных атомов гелия и определяется только термодинамической температурой системы (в отличие от случая $T>T_0$, см. ур. (3)):

$$\mu(N_{nc})_{T \leq T_0} = E_{F_L}(N_{nc})_{T \leq T_0} = -\varphi_{nc} = \frac{N_L \cdot k_B T}{N_{nc}} = k_B T \tag{14}$$

Температура $T_\lambda$ перехода системы в когерентное состояние отвечает температуре завершения перехода всех $Lp$-частиц системы в конденсатное состояние дублетов. Для оценки величины $T_\lambda$ предположим, что в области температур $T<T_0$ эффективная температура фермионов надконденсатной фракции остается равной нулю (как это имело место в области температур $T_{cr}>T \geq T_0$). В таком представлении связь между температурой и числом надконденсатных частиц имеет вид аналогичный выражению (5):

$$E_{F_L}(N_{nc})_{T \leq T_0} = k_B T_{T \leq T_0} = \frac{\hbar^2 k_{nc}^2}{2M_{He}} = \frac{\hbar^2}{2M_{He}} \left( \frac{3\pi^2 N_{nc}}{V} \right)^{2/3} \tag{15}$$

Полагая объемы в соотношениях (15) и (5) одинаковыми $V=V_0=const$, и $N_{nc}(T_\lambda)=N_{0s}$, получаем следующую оценку температуры перехода системы в когерентное состояние:

$$T'_\lambda = T_0 (N_{0s}/N_0)^{2/3} = 2.197 K \tag{16}$$

Полученную оценку мы обозначили, как $T'_\lambda$. Отклонение величины $T'_\lambda$ от экспериментального значения температуры перехода жидкого гелия-4 в когерентное состояние составляет около 1% ($T_\lambda$=2.175±0.005$K$, см., например, [8]). Для определения точного значения температуры фазового перехода жидкого гелия из состояния "гелий-I" в состояние "гелий-II" следует получить явное выражение связи между числом надконденсатных частиц и температурой системы. Получить такое соотношение во всем интервале температур $T \leq T_0$ в аналитическом виде не удается. Однако в области температур $T \leq T'_\lambda$ искомое соотношения можно с хорошей точностью получить в явном виде[7]:

$$T = \alpha_0 T_0 (N_{nc}/N_0)^{2/3} \tag{17a}$$

---

[7] В настоящей работе вывод выражения (17$a$) опущен для сокращения объема статьи. Соответствующие выкладки предполагается привести в другой работе.

где $\alpha_0 \approx 0.989$. Подставляя значения $\alpha_0$ и $N_{0s}/N_0$ в формулу (17*a*) получаем температуру $T_\lambda$, которая практически точно совпадает с экспериментальной температурой перехода гелия-4 в когерентное состояние "гелий-II":

$$T_\lambda = \alpha_0 T_0 \left(\frac{N_{0s}}{N_0}\right)^{2/3} = \alpha_0 T'_\lambda = 2.174 K \qquad (17b)$$

### 1.5 Результаты первой части работы

В первой части работы сформулированы основные идеи и принципы микроскопической физики макроскопического ансамбля атомов гелия. На этой основе построена эволюционная структурно-динамическая модель жидкого гелия. В рамках этой модели определены условия, атомно-молекулярные формы и характеристические термодинамические параметры переходов системы из газового в жидкое состояние, затем в бозе-конденсатное и, наконец, в упорядоченное когерентное состояние. И взаимосвязь между этими параметрами также установлена.

Полученные результаты расчетов температур фазовых переходов первого и второго рода ($T_{cr}$=5.22*K* и $T_\lambda$=2.174*K* соответственно) указывают на то, что идеи, на которых в данной работе строится микроскопическая физика жидкого гелия, адекватно отражают физические процессы, происходящие в гелии-4 в диапазоне температур существования жидкой фазы. В том числе и при его переходе в когерентное состояние с дальним порядком в точке $T_\lambda$.

Изложенные выше представления о микроскопической природе и механизмах взаимодействия атомных частиц [4]He используются во второй части работы для расчета теплоемкости фаз "гелий-I" и "гелий-II". В настоящей работе мы ограничимся расчетами теплоемкости жидкого гелия только в окрестности точки $\lambda$-перехода системы в когерентное состояние.

**Part 2. Теплоемкость жидкого гелия-4 вблизи точки λ-перехода.**

> *"…поведение теплоемкости жидкого гелия вблизи λ-точки описывается эмпирической формулой*
> 
> $$C_V \approx \begin{cases} a + b\ln|T - T_\lambda|, & T < T_\lambda \\ a' + b\ln|T - T_\lambda|, & T > T_\lambda \end{cases}$$
> 
> *Объяснить это явление мы предоставляем читателю в качестве самостоятельного упражнения. Если вам это удастся, опубликуйте!"* ☺
> 
> ***Р. Фейнман, курс лекций "Статистическая Механика", с.44, М., "Мир", 1975***

### 2.1 Метод расчета теплоемкости жидкого гелия при температуре ниже температуры $T_0$ перхода гиротонных квазичастиц в бозе-конденсатное состояние

С позиции квазичастичного представления жидкий гелий при температуре $T=T_0$ является двухкомпонентной системой бигиротонных псевдоатомов, образованной $N_{0s}$ и $N_{0p}$ гиротонными квазичастицами s- и p-типов соответственно. Эти частицы при $T<T_0$ распределяются между фракцией надконденсатных парно-коррелированных фермионных квазичастиц (бигиротонов) и фракцией бозе-конденсатных атомно-молекулярных квазичастиц. При текущей температуре $T$ надконденсатная и конденсатная фракции занимают объемы пропорциональные числу образующих эти фракции частиц $N_{nc}$ и $N_c$ соответственно:

$$V_{nc} = \upsilon N_{nc}; \ V_c = \upsilon N_c; \ V = V_{nc} + V_c; \ N_0 = N_{nc} + N_c \quad (18)$$

где $\upsilon$ - удельный объем жидкости.

Для расчета теплоемкости гелия в зависимости от температуры будем находить энтропии надконденсатной и конденсатной фракций $S_{nc}$ и $S_c$ соответственно. А также энтропию $S_{cnc}$ смешения их частиц. Полная энтропия квазичастичной системы равна:

$$S = S_{nc} + S_c + S_{cnc} \quad (19)$$

Энтропии надконденсатной и конденсатной фракций определяются рассмотренными в первой части работы процессами, формирующими фермионные парно-коррелированные гиротоны и атомно-молекулярную квазичастичную структуру их композитных бозонных состояний. Вклады отдельных составляющих полной энтропии могут быть представлены как явные функции числа надконденсатных квазичастиц $N_{nc}$. Результирующую теплоемкость системы можно представить в виде суммы производных отдельных членов энтропии по числу надконденсатных гиротонных квазичастиц. Вычисления теплоемкости $N_0$-частичного макроансамбля будут выполняться в расчете на одну атомную частицу. Используя зависимость (17а) числа надконденсатных частиц $N_{nc}$ от температуры получаем теплоемкость жидкого гелия, как функцию числа надконденсатных квазичастиц:

$$c = \frac{T}{N_0}\frac{dS}{dT} = \frac{T}{N_0}\sum_i\left(\frac{dS_i}{dT}\right) = \left(\frac{T}{N_0}\frac{dN_{nc}}{dT}\right)\sum_i\left(\frac{dS_i}{dN_{nc}}\right) = \frac{3}{2}\left(\frac{N_{nc}}{N_0}\right)\cdot\sum_i\left(\frac{dS_i}{dN_{nc}}\right) = \sum_i c_i \quad (20)$$

Здесь индекс *i* при символах $S_i$ и $c_i$ нумерует вклады отдельных структурно-динамических подсистем в полные энтропию *S* и теплоемкость *c* жидкого гелия-4.

Для вычисления энтропий подсистем $S_i$ будем определять статистические веса $W_i$ соответствующих состояний: $S_i = \ln W_i$. Согласно такому определению энтропия *S* и теплоемкость *c* в настоящей работе рассчитываются в единицах постоянной Больцмана $k_B$.

Зависимость (17*a*) числа надконденсатных частиц $N_{nc}$ от температуры получена в подраздел 1.4.2 при условии *V=const*. Поэтому определяемая в (20) теплоемкость есть теплоемкость при постоянном объеме ($c=c_V$). В эксперименте теплоемкость $c_s$ жидкого гелия-4 определяется при давлении насыщенных паров (*SVP*). В окрестности температуры $T=T_\lambda$ теплоемкость $c_s$ практически равна теплоемкости $c_p$ при постоянном давлении [9]:

$$c_s \simeq c_P = \frac{1}{N_0}\left[\left(\frac{dE}{dT}\right)_V + P_{SVP}\left(\frac{dV}{dT}\right)_{SVP}\right] = c_V + \Delta c \qquad (21)$$

Поэтому при сравнении теоретических расчетов теплоемкости с экспериментальными данными следует учитывать вклад Δ*c*, который определяется производной объема по температуре при давлении насыщенных паров:

$$\Delta c = \frac{P_{SVP}}{N_0}\left(\frac{dV}{dT}\right)_{SVP} = \frac{T}{N_0}\left(\frac{\partial S}{\partial V}\right)_E\left(\frac{dV}{dT}\right)_{SVP} \qquad (22)$$

Термодинамические величины, относящиеся к области $T>T_\lambda$ (гелий-I) или $T<T_\lambda$ (гелий-II), будем сопровождать нижним индексом "$_+$" или "$_-$" соответственно. Температурную область в непосредственной близости к точке $T=T_\lambda$ будем обозначать как $T \to T_{\lambda+}$ для $T>T_\lambda$ или $T \to T_{\lambda-}$ для $T<T_\lambda$.

## 2.2 Теплоемкость гелия-I.

### 2.2.1 *Структура энтропии $S_+$ гелия-I при температуре ниже температуры бозе-конденсации $T_0$.*

Энтропия и надконденсатной, и конденсатной фракций гелия-I складывается из энтропий, определяемых распределением частиц по псевдоспиновыми степенями свободы гиротонов, и энтропий, определяемых формированием внутренних (псевдоатомных) степеней свободы композитных гиротонных квазичастиц.

Надконденсатная фракция парно-коррелированных состояний гиротонов при $T_0 > T \geq T_\lambda$ образована квазичастицами *Ls*- и *Lp*- компонент. Поэтому энтропию $S_{nc+}$ надконденсатной фракции гелия-I можно представить как сумму двух вкладов. Первый вклад $S_{nc+}^{(sp)}$ определяется распределением частиц надконденсатной фракции по состояниям частиц *Ls*- и *Lp*- компонент. Второй вклад $S_{nc+}^{(b1)}$ обусловлен наличием первичных парных связей между гиротонами: $S_{nc+} = S_{nc+}^{(sp)} + S_{nc+}^{(b1)}$.

В свою очередь, конденсатная дублетная фракция гелия-I является однокомпонентной и формируется на базе квазичастиц только *p*-типа. Поэтому энтропия $S_{c+}$ определяется только процессами образования первичных и вторичных пар. Соответственно, энтропию образования дублетных структурных единиц конденсатной фракции можно представить как сумму энтропии $S_{c+}^{(b1)}$ первичного и $S_{c+}^{(b2)}$ вторичного спаривания: $S_{c+} = S_{c+}^{(b1)} + S_{c+}^{(b2)}$. Таким образом, в соответствии с (19), полная энтропия жидкого гелия-I при температуре $T<T_0$ может быть представлена суммой вкладов:

$$S_+ = S_{cnc+} + S_{nc+}^{(sp)} + S_{nc+}^{(b1)} + S_{c+}^{(b1)} + S_{c+}^{(b2)} \qquad (23)$$

### 2.2.2 Теплоемкость гелия-I при давлении насыщенных паров.

Теплоемкость гелия-I при давлении насыщенных паров включает составляющую $\Delta c = \Delta c_+$ (см. ур. (22)), которая определяется производной энтропии $S_+$ (ур. (23)) по объему. При этом внутренние (псевдоатомные) степени свободы композитных квазичастиц не вносят вклад в давление. И, следовательно, соответствующие составляющие $S_{nc+}^{(b1)}$, $S_{c+}^{(b1)}$ и $S_{c+}^{(b2)}$ полной энтропии $S_+$ не вносят вклад в член $\Delta c_+$ теплоемкости $c_s$. Составляющая $S_{cnc+}$ полной энтропии $S_+$ определяется только распределением частиц конденсатной фракции среди частиц надконденсатной фракции при постоянным числе частиц системы $N_0$ и не зависит явно от объема. Поэтому энтропия $S_{cnc+}$ также не вносит вклад в составляющую теплоемкости $\Delta c_+$. Таким образом, вклад в $\Delta c_+$ вносит только составляющая $S_{nc+}^{(sp)}$ энтропии $S_+$ (см. (23)). Энтропия $S_{nc+}^{(sp)}$ определяется распределением частиц надконденсатной фракции по псевдоспиновым состояниям $Ls$- и $Lp$-псевдофермионов в соответствующем этой подсистеме переменном объеме $V_{nc} = \upsilon N_{nc}$ (см. (18)).

Составляющую $\Delta c_+$ теплоемкости гелия-I (ур. (22)) можно представить в виде:

$$\Delta c_+ = \frac{T}{N_0}\frac{dS_{nc+}^{(sp)}}{dN_{nc}}\frac{dN_{nc}}{dV_{nc}}\left(\frac{N_0 d\upsilon_+}{dN_{nc}}\frac{dN_{nc}}{dT}\right)_{SVP} = \left(\frac{T}{N_0}\frac{dN_{nc}}{dT}\frac{dS_{nc+}^{(sp)}}{dN_{nc}}\right)\left(\frac{N_0 d\upsilon_+}{\upsilon dN_{nc}}\right)_{SVP} = c_{nc+}^{(sp)}\delta_+ \qquad (24)$$

Первый сомножитель $c_{nc+}^{(sp)}$ в правой части уравнения (24) определяет вклад в теплоемкость, связанный с процессом распределения надконденсатных частиц по состояниям компонент $Ls$- и $Lp$- типа при постоянном объеме (см. ниже пункт $b$ подраздела 2.2.3, ф-ла (30)). Второй сомножитель $\delta_+(T) = \left(\frac{N_0 d\upsilon_+}{\upsilon dN_{nc}}\right)_{SVP}$ определяется уменьшением объема гелия-I в результате процесса бозе-конденсации фермионных $Lp$-квазичастиц.

Таким образом, теплоемкость гелия-I при давлении насыщенных паров $c_{s+}$ определяется суммой вкладов:

$$c_{s+}(T) \simeq \left(\frac{T}{N_0}\frac{dN_{nc}}{dT}\right)\cdot\left(\frac{dS_{cnc}}{dN_{nc}} + \frac{dS_{nc+}^{(sp)}}{dN_{nc}} + \frac{dS_{nc+}^{(b1)}}{dN_{nc}} + \frac{dS_{c+}^{(b1)}}{dN_{nc}} + \frac{dS_{c+}^{(b2)}}{dN_{nc}}\right) + \Delta c_+ = \\ = c_{cnc+} + c_{nc+}^{(b1)} + c_{c+}^{(b1)} + c_{c+}^{(b2)} + c_{nc+}^{(sp)}\cdot\gamma_+(T) \qquad (25)$$

Здесь коэффициент $\gamma_+$ при вкладе $c_{nc+}^{(sp)}$ равен:

$$\gamma_+ = 1 + \delta_+ = 1 + \frac{N_0}{\upsilon}\frac{d\upsilon_+}{dN_{nc}} \qquad (26)$$

### 2.2.3 Статистический вес, энтропия и теплоемкость процессов формирования структуры квазичастичных подсистем гелия-I вблизи λ-точки.

#### a. Смешение надконденсатной и конденсатной фракций гелия-I.

Для определения энтропии $S_{cnc+}$ смешения частиц надконденсатной и конденсатной фракций гелия-I в окрестности λ-точки можно записать выражение:

$$S_{cnc+} = \ln W_{cnc+} = \ln \frac{N_0!}{(N_0 - N_{nc})! N_{nc}!} \tag{27}$$

Здесь $W_{cnc+}$ есть статистический вес смешения частиц надконденсатной и конденсатной фракций гелия-I.

В соответствии с ур. (20) из ф-лы (27) получаем, используя формулу Стирлинга, вклад процесса смешения конденсатной и надконденсатной фракций в теплоемкость гелия-I в окрестности $\lambda$-точки:

$$c_{cnc+} = \frac{T}{N_0} \frac{dN_{nc}}{dT} \cdot \frac{dS_{cnc+}}{dN_{nc}} = \left(\frac{3}{2} \frac{N_{nc}}{N_0}\right) \cdot \ln\left(\frac{N_0 - N_{nc}}{N_{nc}}\right) \tag{28}$$

Все нижеследующие расчеты отдельных вкладов в энтропию и теплоемкость выполнены по аналогичной схеме.

### b. Распределение гиротонов надконденсатной фракции гелия-I по Ls- и Lp-состояниям.

Статистический вес $W_{nc+}^{(sp)}$, энтропия $S_{nc+}^{(sp)}$ и теплоемкость $c_{nc+}^{(sp)}$ процесса распределения $N_{nc}$ надконденсатных частиц по состояниям Ls- и Lp-типа вблизи $\lambda$-точки определяются соотношениями:

$$S_{nc+}^{(sp)} = \ln W_{nc+}^{(sp)} = \ln \frac{N_{nc}!}{(N_{nc} - N_{0s})! N_{0s}!} \tag{29}$$

$$c_{nc+}^{(sp)} = \frac{T}{N_0} \frac{dN_{nc}}{dT} \cdot \frac{dS_{nc+}^{(sp)}}{dN_{nc}} = \left(\frac{3}{2} \frac{N_{nc}}{N_0}\right) \cdot \ln\left(\frac{N_{nc}}{N_{nc} - N_{0s}}\right) \tag{30}$$

Как видно, вклад (30) в теплоемкость гелия-I при $N_{nc} \to N_{0s}$ и, следовательно, при $T \to T_{\lambda+}$, стремится к бесконечности.

### c. Вклад формирования первичных пар гиротонов надконденсатной фракции в теплоемкость гелия-I.

Образование первичных псевдофермионных пар определяется спариванием $K$-псевдоэлектронов. При этом, степень вырождения псевдофермионных состояний первичных пар гиротонов определяется степенью спинового вырождения состояний $g_s=2$ $L$-псевдоэлектронов, порождаемых локализованными на атомах гелия вторичными наведенными спинами. Практически все $L$-псевдоэлектроны надконденсатной фракции бигиротонных псевдоатомов при $T \to T_{\lambda+}$ относятся к квазичастицам $s$-типа. В таком случае, вклад внутренних степеней свободы надконденсатной фракции в энтропию $S_{nc+}^{(b1)}$ и теплоемкость $c_{nc+}^{(b1)}$ гелия-I в окрестности $\lambda$-точки находится следующим образом:

$$S_{nc+}^{(b1)} = \ln W_{nc+}^{(b1)} = \ln\left\{\frac{(N_{nc})!}{[(N_{nc}/2)!]^2}\right\}^{g_s} \approx 2N_{nc} \ln 2 \tag{31}$$

$$c_{nc+}^{(b1)} = \left(\frac{T}{N_0} \frac{dN_{nc}}{dT}\right) \cdot \frac{dS_{nc+}^{(b1)}}{dN_{nc}} = \frac{3}{2}\left(\frac{N_{nc}}{N_0}\right) \cdot \ln 4 \tag{32}$$

### d. Вклад внутренних степеней свободы конденсатной фракции в теплоемкость гелия-I.

Формирование дублетных квазичастиц конденсатной фракции определяется процессами образования как первичных, так и вторичных гиротонных пар (см. подраздел 1.3.1). Процесс образования первичных псевдофермионных пар протекает во всем ансамбле $N_0$ атомных частиц путем спаривания $K$-псевдоэлектронов. Поэтому, учитывая вклад первичного спаривания в энтропию и теплоемкость надконденсатной фракции, мы должны также учесть вклад первичного спаривания в энтропию и теплоемкость конденсатной фракции гелия. Первичное спаривание включает в себя процесс локализации на атомах гелия-I вторичных псевдоспинов, который обусловливает "сохранение" фермионных свойств индивидуальных гиротонов. Таким образом, при расчете энтропии $S_{c+}^{(b1)}$ и теплоемкости $c_{c+}^{(b1)}$ мы должны, как и в случае надконденсатных квазичастиц, учесть степень спинового вырождения первичных пар, предваряющую их переход в конденсатное состояние путем вторичного спаривания. Эти вклады определяются аналогично соответствующим вкладам надконденсатной фракции при $T \to T_{\lambda+}$ (см. ур. (31, 32)):

$$S_{c+}^{(b1)} = \ln W_{c+}^{(b1)} = \ln \left\{ \frac{(N_c)!}{[(N_c/2)!]^2} \right\}^{g_s} \approx 2 N_c \ln 2 \qquad (33)$$

$$c_{c+}^{(b1)} = \left( \frac{T}{N_0} \frac{dN_{nc}}{dT} \right) \cdot \left( \frac{d\left(S_{c+}^{b1}\right)}{dN_{nc}} \right) = -\frac{3}{2} \left( \frac{N_{nc}}{N_0} \right) \cdot \ln 4 \qquad (34)$$

Здесь $g_s = 2$ есть спиновая степень вырождения состояний первичных пар, которая определяется локализованными на атомах не спаренными вторичными псевдоспинами $s=1/2$.

Собственно переход гиротонов в бозе-конденсатное состояние связан, согласно подразделу 1.3.1, с процессом вторичного спаривания парно-коррелированных квазичастиц. Физическое содержание данного процесса обусловлено спариванием первичных пар путем формирования общих $Lp$ орбиталей $L$-псевдоэлектронов дублетной псевдомолекулярной квазичастицы. В результате псевдоспиновые состояния четырех гиротонов перестают быть независимыми.

Состояния бигиротонных дублетов определяются суперпозицией состояний $s$- и $p$-псевдоэлектронных пар (см. подраздел 1.3.1) . С этой точки зрения эффект формирования дублетов можно условно представить как совокупный результат двух процессов. Во-первых, новые связи образуются путем $p$- спаривания $Lp$-псевдоэлектронов., входящих в состав двух соседних первичных пар Степень вырождения $p$-парных состояний равна $g_1 = 3$. И, во-вторых, новые связи образуются также и между гиротонами первичной пары путем $s$- спаривания их $Lp$-псевдоэлектронов. При этом степень вырождения псевдоспинового состояния первичной пары $g_s = 2$ уменьшается до величины степени вырождения $s$- парного состояния $L$-псевдоэлектронов $g_0 = 1$. С учетом этих обстоятельств результирующая степень вырождения парных состояний гиротонов, образуемых в процессе вторичного спаривания, будет равна $g_a = g_1 + g_0 - g_s = 2$. В таком случае, для определения энтропии $S_{c+}^{(b2)}$ и теплоемкости $c_{c+}^{(b2)}$ процесса образования дублетных состояний конденсатных бигиротонов получаем выражения:

$$S_{c+}^{(b2)} = \ln W_{c+}^{(b2)} = \ln\left\{\frac{(N_c)!}{\left[(N_c/2)!\right]^2}\right\}^{g_a} \approx 2N_c \ln 2 \qquad (35)$$

$$c_{c+}^{(b2)} = \left(\frac{T}{N_0}\frac{dN_{nc}}{dT}\right)\cdot\left(\frac{d\left(S_{c+}^{b2}\right)}{dN_{nc}}\right) = -\frac{3}{2}\left(\frac{N_{nc}}{N_0}\right)\cdot\ln 4 \qquad (36)$$

### 2.2.4 *Температурная зависимость теплоемкости гелия-I вблизи λ-точки.*

Согласно результатам подразделов 2.2.2 и 2.2.3, теплоемкость гелия-I при давлении насыщенных паров вблизи температуры $T_{\lambda+}$ определяется уравнением:

$$c_{T>T_\lambda} = c_{\tau+} = \frac{3}{2}\frac{N_{nc}}{N_0}\cdot\left[\gamma_+\cdot\ln\left(\frac{N_{nc}}{N_{nc}-N_{0s}}\right) + \ln\left(\frac{N_0-N_{nc}}{N_{nc}}\right) - \ln 4\right] \qquad (37)$$

Используя соотношение (17*a*) получаем температурную зависимость теплоемкости гелия-I в области температуры λ-перехода, которую можно представить в виде линейной функции логарифма переменной $\tau_+$:

$$c_{\tau+} = a_+ + b_+\ln(\tau_+), \qquad (38)$$

$$\tau_+ = |(T-T_\lambda)/T_\lambda| = (T/T_\lambda - 1); \qquad (39)$$

коэффициенты $a_+$ и $b_+$ уравнения (38) равны, соответственно:

$$a_+ = \frac{3}{2}\frac{N_{0s}}{N_0}\cdot\left[\ln\left(\frac{1}{4}\cdot\frac{N_{0p}}{N_{0s}}\right) - \gamma_+(T_\lambda)\cdot\ln\frac{3}{2}\right], \qquad (40a)$$

$$b_+(T_\lambda) = -\gamma_+(T_\lambda)\cdot\frac{3}{2}\cdot\frac{N_{0s}}{N_0}; \qquad (40b)$$

значения величин $N_{0s}/N_0$ и $N_{0p}/N_{0s}$ определены соотношениями (7).

Таким образом, на базе сформулированных в первой части работы принципов микроскопической физики макроансамбля атомов гелия-4 получена теоретическая зависимость теплоемкости от температуры по форме совпадающая с эмпирической зависимостью, описывающей соответствующие экспериментальные данные гелия-I (см. эпиграф на стр. 17).

### 2.3 Теплоемкость гелия-II.

#### 2.3.1 *Структура энтропии $S_-$ гелия-II.*

Общая схема расчета теплоемкости гелия-II такая же, как и для гелия-I. Специфика расчета энтропии жидкого гелия-II связана с переходом системы в состояние с дальним порядком. Такое состояние гелия-II определяется формированием единой для всей системы пространственно-временной структуры внутреннего вихревого поля, которое обуславливает взаимозависимое распределение гиротонных квазичастиц по состояниям *s*- и *p*- типов.

Надконденсатная и конденсатная фракции в λ-точке состоят только из частиц *Ls*- и *Lp*- типа соответственно: $N_{nc}(T_\lambda)=N_{0s}$; $N_c(T_\lambda)=N_{0p}$. Степень упорядоченности системы в этой точке характеризуется параметром порядка $\chi=0$. Энтропия смешения частиц надконденсатной и конденсатной фракций $S_{cnc}(T_\lambda)=S_{cnc}^{(\lambda)}$ в этом случае соответствует энтропии $S^{(sp)}(T_\lambda)$ распределения всех надконденсатных $N_{0s}$ частиц $Ls$- типа среди всех конденсатных $N_{0p}$ частиц $Lp$- типа.

При понижении температуры увеличение числа конденсатных частиц в гелии-II происходит за счет $N_{sc}=(N_{0s}-N_{nc})=(N_c-N_{0p})$ парно-коррелированных квазичастиц $Ls$-типа, которые переходят в конденсатное состояние при участии $Lp$-частиц конденсатных дублетов. В результате в конденсатной фракции гелия-II образуется упорядоченная подсистема $N_{b3}=3(N_c-N_{0p})$ гиротонных квазичастиц в форме бигиротонных триплетов (см. подраздел 1.4.1).

При температуре $T<T_\lambda$ бозонизация $N_{sc}(T)$ квазичастиц $Ls$-типа и формирование на их основе упорядоченной подсистемы конденсатной фракции исключает эти $Ls$-частицы из процесса смешения $N_{nc}(T_\lambda)=N_{0s}$ и $N_c(T_\lambda)=N_{0p}$ частиц надконденсатной и конденсатной фракций, соответственно, имевшего место при $T=T_\lambda$ и $\chi=0$. Поэтому, для вычисления энтропии $S_{cnc-}(T)$ смешения конденсатной и надконденсатной фракций гелия-II с ненулевым параметром порядка ($\chi \neq 0$) необходимо вычесть энтропию $S_{c-}^{(sp)}(T)$ смешения $N_{sc}(T)=N_{0s}-N_{nc}(T)$ и $N_{0p}$ конденсатных частиц $Ls$- и $Lp$-типа, соответственно, из энтропии $S_{cnc}(T_\lambda)=S^{(sp)}(T_\lambda)$, отвечающей смешению всех $Ls$-частиц среди всех $Lp$-частиц в состоянии системы с нулевым параметром порядка ($\chi=0$):

$$S_{cnc-}(T) = S_{cnc}(T_\lambda) - S_{c-}^{(sp)}(T) = S_{cnc}^{(\lambda)} - S_{c-}^{(sp)} \tag{41}$$

Таким образом, общая структура энтропии гелия-II, в соответствии с изложенным соображениям и согласно (19, 41), может быть записана в виде:

$$S_- = S_{cnc}^{(\lambda)} - S_{c-}^{(sp)} + S_{nc-} + S_{c-} \tag{42a}$$

Ниже температуры $\lambda$-перехода надконденсатная фракция парных состояний гиротонов включает только частицы $Ls$-компоненты. Соответственно, энтропия надконденсатной фракции $S_{nc-}$ определяется только формированием внутренних степеней свободы бигиротонных псевдоатомов: $S_{nc-}=S_{nc-}^{(b1)}$. Данный процесс установления первичных парных состояний идет между всеми $N_0$ частицами системы. Следовательно, первичное спаривание также вносит вклад $S_{c-}^{(b1)}$ в полную энтропию $S_{c-}$ конденсатной фракции гелия-II. Кроме того, энтропия $S_{c-}$ включает вклады $S_{c-}^{(b2)}$ и $S_{c-}^{(b3)}$, которые определяются внутренними степенями свободы неупорядоченной и упорядоченной подсистем дублетных и триплетных квазичастиц соответственно (см. подраздел 1.4.1).

Конденсатная неупорядоченная подсистема дублетов фазы "гелий-II" состоит из $N_{b2}=N_c-3(N_c-N_{0p})$ гиротонных квазичастиц $Lp$- компоненты и, соответственно, не вносит в полную энтропию вклад, связанный с распределением частиц по состояниям различных компонент. Упорядоченная подсистема триплетов конденсатной фракции формируется на основе квазичастиц $Ls$- и $Lp$- компонент. Однако данная подсистема, в силу ее упорядоченного состояния, также не вносит связанный с распределением $Ls$- и $Lp$- частиц относительно друг друга вклад в энтропию $S_{c-}$ конденсатной фракции гелия-II. Собственно вклад процесса упорядочения в энтропию системы включен в энтропию $S_{cnc-}(T)$ смешения частиц конденсатной и надконденсатной фракций в форме синтропийной составляющей $-S_{c-}^{(sp)}$ (ф-ла (41)).

Значения энтропий $S_{c-}^{(b2)}$ и $S_{c-}^{(b3)}$ двух подсистем конденсатной фракции гелия-II конечны, а их вклад в энтропию конденсатной фракции прямо пропорционален числу образующих эти подсистемы гиротонных квазичастиц. Число частиц $N_{b3}=3(N_c-N_{0p})$, формирующих псевдомолекулярные триплеты, стремится к нулю при $T \to T_{\lambda-}$. И, соответственно, число частиц $N_{b2}=N_c-3(N_c-N_{0p})$, формирующих дублеты, стремится к $N_c$. Поэтому, практически, в области $\lambda$-точки энтропия внутренних степеней свободы конденсатной фракции гелия-II определяется только суммой членов $S_{c-}^{(b1)}$ и $S_{c-}^{(b2)}$. Следовательно, энтропия гелия-II в области температуры $\lambda$-перехода может быть рассчитана как сумма вкладов:

$$S_- \simeq S_{cnc}^{(\lambda)} - S_{c-}^{(sp)} + S_{nc-}^{(b1)} + S_{c-}^{(b1)} + S_{c-}^{(b2)} \tag{42b}$$

### 2.3.2 Теплоемкость гелия-II при давлении насыщенных паров.

По тем же основаниям, что и для гелия-I (см. подраздел 2.2.2), только одна составляющая $-S_{c-}^{(sp)}$ энтропии $S_-$ (см. (42b)) вносит вклад в член $\Delta c = \Delta c_-$ теплоемкости $c_s$ гелия-II (см. ур. (21, 22)). Согласно подразделу 2.3.1, эта составляющая отражает эффект упорядочения частиц конденсатной фракции по состояниям $s$- и $p$- типов в соответствующем этой подсистеме переменном объеме $V_c = \upsilon N_c$ (см. (18)).

Таким образом, принимая во внимание вышеизложенное и осуществляя в ур. (22) преобразования, аналогичные преобразованиям при выводе ф-лы (24) в подразделе 2.2.2, получаем выражение для составляющей $\Delta c_-$ атомной теплоемкости $c_s$ гелия-II в виде:

$$\Delta c_- = \frac{P_{SVP}}{N_0}\left(\frac{dV_-}{dT}\right)_{SVP} = \left(\frac{T}{N_0}\frac{dN_{nc}}{dT}\frac{d(-S_{c-}^{(sp)})}{dN_{nc}}\right)\left(\frac{-N_0 d\upsilon_-}{\upsilon dN_{nc}}\right)_{SVP} = c_{c-}^{(sp)} \cdot \delta_- \tag{43}$$

Здесь первый сомножитель $c_{c-}^{(sp)}$ в правой части уравнения (43) определяет вклад в теплоемкость при постоянном объеме, связанный с процессом упорядочения переходящих в конденсатное состояние надконденсатных частиц $Ls$- компоненты относительно ранее перешедших в конденсатное состояние частиц $Lp$- компоненты (см. ниже, пункт **b** подраздела 2.3.3, ф-ла (48)). Второй сомножитель в правой части уравнения (43) $\delta_-(T) = \left(\frac{-N_0 d\upsilon_-}{\upsilon dN_{nc}}\right)_{SVP}$ определяется увеличением удельного объема гелия-II в процессе бозе-конденсации $Ls$- фермионных квазичастиц. Таким образом, в соответствии с формулами (19-23) и (41, 42b, 43), полная теплоемкость гелия-II при давлении насыщенных паров в области точки $\lambda$-перехода определяется суммой вкладов:

$$c_{s-}(T) \simeq \left(\frac{T}{N_0}\frac{dN_{nc}}{dT}\right) \cdot \left(\frac{d(-S_{c-}^{(sp)})}{dN_{nc}} + \frac{dS_{nc-}^{(b1)}}{dN_{nc}} + \frac{dS_{c-}^{(b1)}}{dN_{nc}} + \frac{dS_{c-}^{(b2)}}{dN_{nc}}\right) + \Delta c_- =$$
$$= c_{nc-}^{(b1)} + c_{c-}^{(b1)} + c_{c-}^{(b2)} + \gamma_-(T) \cdot c_{c-}^{(sp)} \tag{44}$$

$$\gamma(T<T_\lambda) = \gamma_- = (1+\delta_-) = 1 - \frac{N_0}{\upsilon}\frac{d\upsilon_-}{dN_{nc}} \tag{45}$$

### 2.3.3 Статистический вес, энтропия и теплоемкость процессов формирования структуры квазичастичных подсистем гелия-II вблизи λ-точки.

#### a. Смешение надконденсатной и конденсатной фракций в λ-точке.

Согласно подразделу 2.3.1, энтропия смешения конденсатной и надконденсатной фракций гелия-II $S_{cnc-}(T)$ с отличным от нуля параметром порядка $\chi$ определяется разностью двух вкладов: $S_{cnc-}(T) = S_{cnc}(T_\lambda) - S_{c-}^{(sp)}$ (см. ф-лу (41)). Первая составляющая $S_{cnc}(T_\lambda)$ энтропии смешения $S_{cnc-}(T)$ численно совпадает с энтропией $S^{(sp)}(T_\lambda)$:

$$S_{cnc}^{(\lambda)} = \ln W_{cnc}^{(\lambda)} = \ln\left(\frac{N_0!}{N_{0s}!N_{0p}!}\right) \tag{46}$$

Эта составляющая не зависит от температуры и не вносит вклад в теплоемкость гелия-II. В соответствии с изложенными в подразделе 2.3.1 соображениями, второй вклад $S_{c-}^{(sp)}$ исключается из первого вклада в силу перехода $N_{sc}(T)=N_{0s}-N_{nc}(T)$ надконденсатных $Ls$-частиц в упорядоченное состояние относительно $Lp$-частиц. Таким образом, температурная зависимость энтропии смешения конденсатной и надконденсатной фракций при $T<T_\lambda$ определяется процессом формирования упорядоченной подсистемы конденсатной фракции гелия-II. Синтропия $-S_{c-}^{(sp)}$ данного процесса и соответствующий вклад в теплоемкость гелия-II приведены в следующем пункте подраздела.

#### b. Вклад процесса упорядочения гиротонных квазичастиц двухкомпонентной системы конденсатной фракции в теплоемкость гелия-II.

Статистический вес $W_{c-}^{(sp)}$ и энтропия $S_{c-}^{(sp)} = \ln W_{c-}^{(sp)}$ распределения $N_{sc}(T)=N_{0s}-N_{nc}(T)$ квазичастиц $Ls$-типа среди $N_{0p}$ квазичастиц $Lp$-типа конденсатной фракции определяются соотношением:

$$S_{c-}^{(sp)} = \ln\frac{N_c!}{N_{0p}!(N_c - N_{0p})!} \tag{47}$$

Таким образом, согласно (41) теплоемкость процесса смешения надконденсатной и конденсатной фракций гелия при $T<T_\lambda$ определяется производной по температуре синтропии упорядочения двухкомпонентной системы $-S_{c-}^{(sp)}$:

$$c_{c-}^{(sp)} = \left(\frac{T}{N_0}\frac{dN_{nc}}{dT}\right)\cdot\frac{d\left(-S_{c-}^{(sp)}\right)}{dN_{nc}} = \left(\frac{3}{2}\frac{N_{nc}}{N_0}\right)\ln\left(\frac{N_c}{N_c - N_{0p}}\right) \tag{48}$$

Вклад $c_{c-}^{(sp)}$ в теплоемкость гелия-II при $N_{nc} \to N_{0s}$ и, следовательно, при $T \to T_{\lambda-}$ стремится к бесконечности (как и в случае теплоемкости $c_{nc+}^{(sp)}$ гелия-I при $T \to T_{\lambda+}$, см. формулу (30).

#### c. Вклад формирования первичных пар гиротонов надконденсатной фракции в теплоемкость гелия-II.

Выражения для определения статистического веса $W_{nc-}^{(b1)}$, энтропии $S_{nc-}^{(b1)} = \ln W_{nc-}^{(b1)}$ и теплоемкости процесса формирования первичных пар фермионных гиротонов надконденсатной фракции в окрестности температуры $T_\lambda$ для гелия-II совпадают с выражениями для гелия-I (формулы (31) и (32)):

$$S_{nc-}^{(b1)} = \ln W_{nc-}^{(b1)} = \ln\left\{\frac{N_{nc}!}{\left[(N_{nc}/2)!\right]^2}\right\}^2 = N_{nc}\ln 4 \qquad (49)$$

$$c_{nc-}^{(b1)} = \left(T\frac{d(N_{nc}/N_0)}{dT}\right)\cdot\frac{dS_{nc-}^{(b1)}}{dN_{nc}} = \left(\frac{3}{2}\frac{N_{nc}}{N_0}\right)\ln 4 \qquad (50)$$

### d. Теплоемкость внутренних степеней свободы композитных квазичастиц конденсатной фракции гелия-II.

В гелии-II, как и в гелии-I, вклад внутренних степеней свободы в энтропию конденсатной фракции вносят процессы формирования как первичных, так и вторичных пар.

Процесс образования первичных фермионных пар не связан и с переходом гелия в когерентное состояние. Как следствие, первичное спаривание вносит вклады в энтропию и теплоемкость конденсатной фракции гелия-II аналогичные вкладам в энтропию и теплоемкость конденсатной фракции гелия-I (см. (33, 34 )):

$$S_{c-}^{(b1)} = \ln W_{c-}^{(b1)} = \ln\left\{\frac{(N_c)!}{\left[(N_c/2)!\right]^2}\right\}^2 \simeq N_c\cdot\ln 4 \qquad (51)$$

$$c_{c-}^{(b1)} = \left(T\frac{d(N_{nc}/N_0)}{dT}\right)\cdot\frac{dS_{c-}^{(b1)}}{dN_{nc}} = -\left(\frac{3}{2}\frac{N_{nc}}{N_0}\right)\cdot\ln 4 \qquad (52)$$

Согласно изложенными в подразделе 1.4.1 представлениям, переход $Ls$-частиц в бозе-конденсатное состояние обусловлен эффектом их вторичного спаривания при посредничестве частиц дублетной конденсатной фракции. Это приводит к разделению $N_c$ частиц дублетной конденсатной фракции гелий-II на две составляющие – упорядоченную подсистему вновь образующихся триплетов и неупорядоченную подсистему ранее образованных дублетных квазичастиц.

Согласно подразделу 2.3.1, в области температур $T\to T_{\lambda-}$ можно пренебречь вкладом внутренних степеней свободы триплетной подсистемы в энтропию и теплоемкость конденсатной фракции гелия-II. В свою очередь, вклад $N_{b2}=N_c-3(N_c-N_{0p})$ частиц дублетной подсистемы в энтропию конденсатной фракции гелия-II, обусловленный эффектом вторичного спаривания, рассчитывается аналогично соответствующему вкладу в энтропию конденсатной фракции гелий-I (см. подраздел 2.2.3, пункт *d*, ф-ла (35)):

$$S_{c-}^{(b2)} = \ln W_{c-}^{(b2)} = \ln\left\{\frac{(3N_{0p}-2N_c)!}{\left\{\left[(3N_{0p}-2N_c)/2\right]!\right\}^2}\right\}^{g_a} \approx (3N_{0p}-2N_c)\ln 4 \qquad (53)$$

Число частиц конденсатной фракции гелия-II при $T\to T_{\lambda-}$ практически полностью определяется числом частиц дублетной подсистемы ($N_{b2}\to N_c$). Поэтому вклад дублетной подсистемы конденсатной фракции в теплоемкость гелия-II в области температуры $T_{\lambda-}$ равен:

$$c_{c-}^{(b2)} = \left( \frac{T}{N_0} \frac{dN_{nc}}{dT} \right) \cdot \frac{dS_{c-}^{(b2)}}{dN_{nc}} = \left( \frac{3}{2} \frac{N_{nc}}{N_0} \right) \cdot 2\ln 4 \qquad (54)$$

Сопоставление вкладов, вносимых фракцией дублетных квазичастиц в теплоемкость гелия-I $c_{c+}^{(b2)}$ и гелия-II $c_{c-}^{(b2)}$ (ф-лы (36) и (54), соответственно), показывает, что эффект разделения конденсатной фракции гелия на упорядоченную и неупорядоченную подсистемы бигиротонов в $\lambda$-точке приводит к конечному скачку теплоемкости $c_s$ равному, согласно формулам (36, 54) и (7), :

$$\delta c_\lambda = c_{c-}^{(b2)} - c_{c+}^{(b2)} = \frac{3}{2}\left(\frac{N_{0s}}{N_0}\right) \cdot 3\ln 4 = 2.411 \qquad (55)$$

### 2.3.4 *Температурная зависимость теплоемкости гелия-II вблизи $\lambda$-точки.*

Согласно результатам подраздела 2.3.3 и в соответствии с (44, 45), теплоемкость гелия-II при давлении насыщенных паров вблизи температуры $T_{\lambda-}$ определяется следующим образом:

$$c_{T<T_\lambda} = c_{\tau-} = \frac{3}{2}\frac{N_{nc}}{N_0}\left[ \gamma_-(T) \cdot \ln\left( \frac{N_c}{N_c - N_{0p}} \right) + 2\ln 4 \right] \qquad (56)$$

Выражение (56) температурной зависимости теплоемкости гелия-II в окрестности температуры $T_\lambda$ приводится к виду аналогичному (38):

$$c_{\tau-} = a_- + b_-\ln(\tau_-), \qquad (57)$$

где переменная $\tau_-$ уравнения (57) определяется выражением:

$$\tau_- = \left|(T-T_\lambda)/T_\lambda\right| = (1 - T/T_\lambda). \qquad (58)$$

Коэффициенты $a_-$ и $b_-$ уравнения (57) равны, соответственно:

$$a_-(T_\lambda) = \frac{3}{2}\frac{N_{0s}}{N_0} \cdot \left[ \ln\left( 16 \cdot \frac{N_{0p}}{N_{0s}} \right) - \gamma_-(T_\lambda) \cdot \ln\frac{3}{2} \right], \qquad (59a)$$

$$b_-(T_\lambda) = -\gamma_-(T_\lambda) \cdot \frac{3}{2}\frac{N_{0s}}{N_0}; \qquad (59b)$$

### 2.4 Сравнение теоретических расчетов теплоемкости жидкого гелия-4 с экспериментальными данными в окрестности точки $\lambda$-перехода системы в когерентное состояние.

Для численного сравнения теоретической и экспериментальной зависимостей теплоемкости от температуры в окрестности $\lambda$-точки необходимо определить значения коэффициентов $\gamma(T_{\lambda+}) = \gamma_+^{(\lambda)}$ и $\gamma(T_{\lambda-}) = \gamma_-^{(\lambda)}$ (ф-лы (26, 45) ). В настоящей работе для этого использовались экспериментальные данные по температурной зависимости плотности жидкого гелия при давлении насыщенных паров (см. работу [8]). Соответствующая оценка коэффициента $\delta(T_{\lambda+}) = \delta_+^{(\lambda)} = (1 - \gamma_+^{(\lambda)})$ при движении к температуре $\lambda$-перехода из области $T>T_\lambda$ дает приближенное значение, равное:

$$\delta_+^{(\lambda)} \approx 0.07 \pm 0.02 \qquad (60a)$$

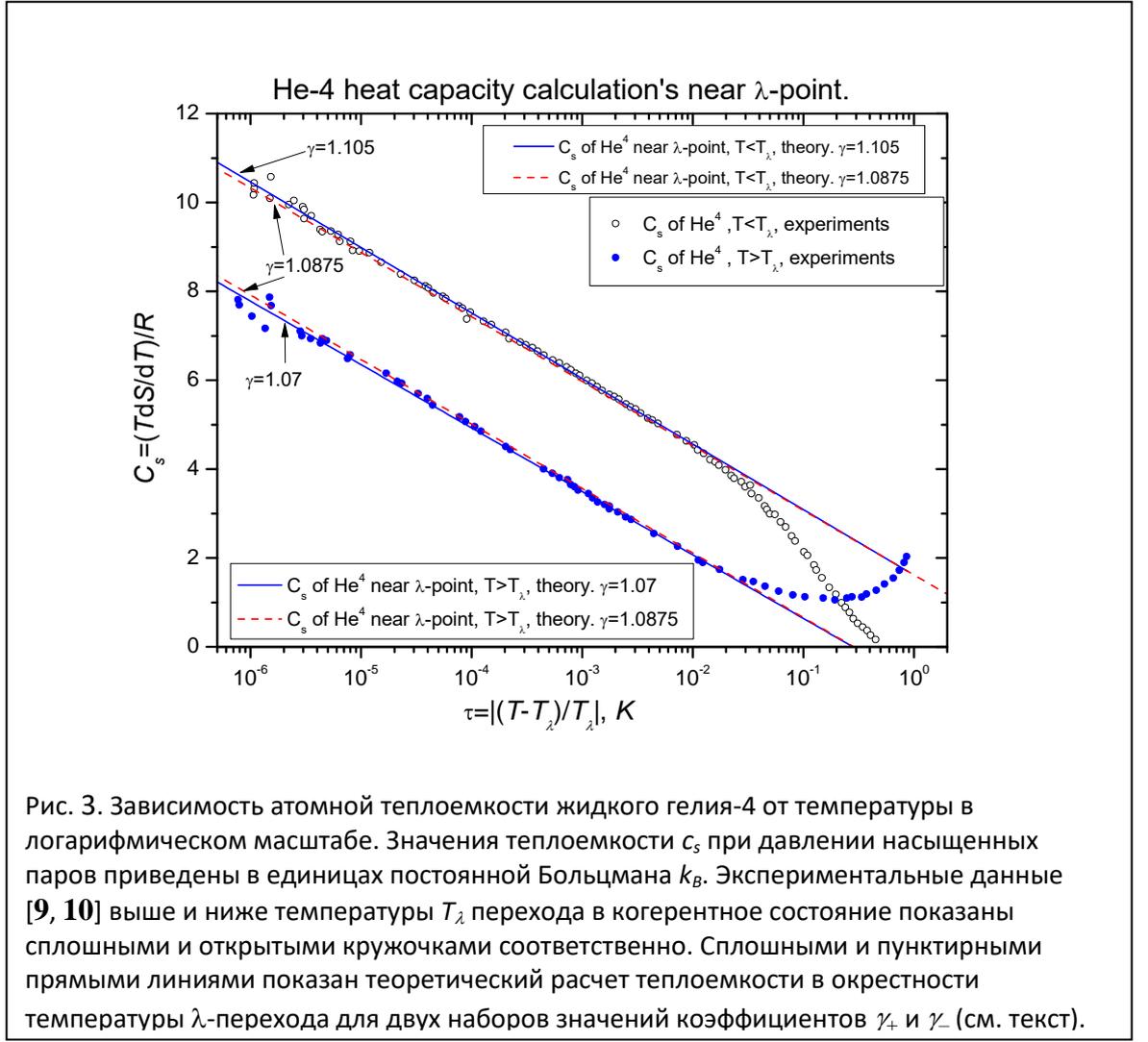

Рис. 3. Зависимость атомной теплоемкости жидкого гелия-4 от температуры в логарифмическом масштабе. Значения теплоемкости $c_s$ при давлении насыщенных паров приведены в единицах постоянной Больцмана $k_B$. Экспериментальные данные [**9, 10**] выше и ниже температуры $T_\lambda$ перехода в когерентное состояние показаны сплошными и открытыми кружочками соответственно. Сплошными и пунктирными прямыми линиями показан теоретический расчет теплоемкости в окрестности температуры $\lambda$-перехода для двух наборов значений коэффициентов $\gamma_+$ и $\gamma_-$ (см. текст).

Оценка коэффициента $\delta(T_{\lambda-}) = \delta_-^{(\lambda)} = \left(1 - \gamma_-^{(\lambda)}\right)$ при движении к температуре $\lambda$-перехода из области $T<T_\lambda$ дает значение:

$$\delta_-^{(\lambda)} \approx 0.105 \pm 0.02 \tag{60b}$$

Подставляя оценочные величины $\gamma_{+/-}^{(\lambda)} = 1 + \delta_{+/-}^{(\lambda)}$ и расчетные значения относительных долей гиротонных квазичастиц $N_{0s}/N_0$ и $N_{0p}/N_0$ s- и p-типов соответственно (см. результаты (7), подраздел 1.2.2) в выражения для коэффициентов $a_+$, $b_+$ и $a_-$, $b_-$ (формулы (40a,b) и (59a,b)), получим следующие соотношения для температурной зависимости атомной теплоемкости гелия-I и гелия-II в области $\lambda$-перехода (в единицах $k_B$):

$$c_{\tau+} = a_+ + b_+ \ln(\tau_+) = -0.7873 - 0.6203 \cdot \ln(\tau_+) \tag{61}$$

$$c_{\tau-} = a_- + b_- \ln(\tau_-) = 1.6155 - 0.6406 \cdot \ln(\tau_-) \tag{62}$$

На рис. 3 приведены результаты прецизионных измерений теплоемкости жидкого гелия при давлении насыщенных паров (см. [9, 10]). На том же рисунке показаны теоретические расчеты теплоемкости гелия-4 в области $\lambda$-перехода (формулы (61) и (62) для $T \rightarrow T_{\lambda+}$ и $T \rightarrow T_{\lambda-}$ соответственно). Как видно, результаты расчетов теплоемкости в области температур $|T - T_\lambda| < 10^{-2} K$ находятся в очень хорошем согласии с экспериментальными данными.

Как следует из (61) и (62), коэффициенты $b_+$ и $b_-$ (ф-лы (40*b*) и (59*b*)) при логарифме $\tau=|T/T_\lambda-1|$ (уравнения (38) и (57)) отличаются примерно на 3%. Точность экспериментальных данных, используемых для определения коэффициентов $\delta_+^{(\lambda)}$ и $\delta_-^{(\lambda)}$, не позволяет однозначно утверждать, что коэффициенты $b_+$ и $b_-$ различны[8]. Поэтому на рис. 3 приводятся также результаты расчета теплоемкости с равными значениями коэффициентов $\gamma=1+\delta$, равными для случаев $T\to T_{\lambda+}$ и $T\to T_{\lambda-}$:

$$\gamma_{+/-}^{(\lambda)} = 1+\delta_{+/-}^{(\lambda)} = 1+\left(\delta_+^{(\lambda)}+\delta_-^{(\lambda)}\right)/2 = 1.0875 \tag{63}$$

Это обусловливает равенство коэффициентов $b_+$ и $b_-$ уравнений (38) and (57) соответственно. В таком случае, эти уравнения принимают вид:

$$\begin{aligned}c_{\tau+} &= a'_+ + b'_+ \ln(\tau_+) = -0.7914 - 0.6305\cdot\ln(\tau_+) \\ c_{\tau-} &= a'_- + b'_- \ln(\tau_-) = 1.6196 - 0.6305\cdot\ln(\tau_-)\end{aligned} \tag{64}$$

Как видно из рис. 3Рис. , результаты теоретических расчетов теплоемкости с коэффициентами $a'_+, b'_+$ и $a'_-, b'_-$ уравнений (64) изменяются незначительно по отношению предыдущим расчетам с коэффициентами $a_+, b_+$ и $a_-, b_-$ уравнений (61) и (62). И, таким образом, согласие между расчетными значениями теплоемкости и экспериментальными данными для случая с одинаковыми коэффициентами $b'_+ = b'_-$ при логарифме приведенной температуры $\tau$ (и, соответственно, с одинаковыми значениями $\gamma_+^{(\lambda)} = \gamma_-^{(\lambda)} = \gamma_{+/-}^{(\lambda)} = 1+\delta_{+/-}^{(\lambda)}$) остается очень хорошим[9].

### 2.5 Результаты второй части работы

Впервые удалось теоретически получить асимптотику теплоемкости вида $c=a+b*ln(|T-T_\lambda|)$ для реальной трехмерной системы в окрестности температуры $T_\lambda$ фазового перехода второго рода. При этом получено отличное количественное согласие расчета с экспериментом . На примере жидкого гелия-4 продемонстрирована физическая состоятельность излагаемых в работе идей для уяснения микроскопической природы кооперативных квантовых явлений.

**Заключение**

В данной статье развивается микроскопический подход к описанию поведения жидкого гелия-4 на основе концепции гироатомных квазичастиц, обладающих псевдоатомными и псевдоспиновыми степенями свободы. Согласно развиваемым в данной работе идеям жидкий гелий представляет собой материал, структурные элементы которого формируются подобно внешним электронным оболочкам атомов и/или молекул.

---

[8] В то же время, можно достаточно уверенно полагать, что искомые значения обоих коэффициентов лежат в пределах указанных в ф-лах (60*a,b*) погрешностей.

[9] Вопрос о детальном ходе температурной зависимости логарифмической сингулярности требует теоретического расчета значения параметров $\gamma_+$ и $\gamma_-$ уравнений (61) и (62). Такой расчет, в принцие, может быть выполнен с позиции изложенных в настоящей работе представлений о микроскопической физике гелия. Решение этой задачи будет предметом отдельной работы.

Сформулированные в настоящей работе физические принципы и механизмы установления и эволюции межчастичных корреляционных связей атомов гелия-4 в процессе понижения температуры могут стать, как представляется, концептуальной основой для понимания микроскопической физики известных кооперативных квантовых эффектов и описания их свойств с единой точки зрения.

## Литература